\documentclass[aps,prx,twocolumn,superscriptaddress]{revtex4}
\usepackage[usenames,dvipsnames]{color}

\usepackage[latin1]{inputenc}
\usepackage{graphicx}
\usepackage{amssymb}
\usepackage{subfigure}
\usepackage{stackrel}
\usepackage{soul}
\usepackage{amsmath}
\usepackage{amsfonts}
\usepackage{bm}
\usepackage{bbold} 
\usepackage{color}
\newcommand{\ket}[1]{\ensuremath{\left|{#1}\right\rangle}}
\newcommand{\bra}[1]{\ensuremath{\left\langle{#1}\right |}}

\newcommand{\bsy}[1]{\ensuremath{\boldsymbol{#1}}}

\newcommand{\bmath}[1]{{\mathbb{#1}}}

\newcommand{\vectg}[1]{\boldsymbol{#1}}

\newcommand{\beq}{\begin{equation}}
\newcommand{\eeq}{\end{equation}}
\newcommand{\bse}{\begin{subequations}}
\newcommand{\ese}{\end{subequations}}
\newcommand{\bea}{\begin{eqnarray}}
\newcommand{\eea}{\end{eqnarray}}
\newcommand{\bit}{\begin{itemize}}
\newcommand{\eit}{\end{itemize}}
\newcommand{\bpmatrix}{\begin{pmatrix}}
\newcommand{\epmatrix}{\end{pmatrix}}

\newcommand{\bfal}{{\vec{t}}}

\newcommand{\bfbe}{{\vec{s}}}

\newcommand{\Mxal}{\bmath M_{x,\vec{t}}}
\newcommand{\Mpal}{\bmath M_{p,\vec{t}}}
\newcommand{\Mpbe}{\bmath M_{p,\vec{s}}}
\newcommand{\Mxbe}{\bmath M_{x,\vec{s}}}
\newcommand{\Mnual}{\bmath M_{\nu,\vec{t}}}

\newcommand{\Mmual}{\bmath M_{\mu,\vec{t}}}

\begin{document}


\title{Systematic construction of genuine multipartite entanglement criteria in continuous variable systems using uncertainty relations}






\author{F. Toscano}
\affiliation{Instituto de F\'{\i}sica, Universidade Federal do Rio de
Janeiro, Caixa Postal 68528, Rio de Janeiro, RJ 21941-972, Brazil}

\author{A. Saboia}
\affiliation{Instituto de F\'{\i}sica, Universidade Federal do Rio de
Janeiro, Caixa Postal 68528, Rio de Janeiro, RJ 21941-972, Brazil}

\author{A. T. Avelar}
\affiliation{Instituto de F\'{\i}sica, Universidade Federal de Goi\'as,
Caixa Postal 131, Goi\^ania, GO 74001-970, Brazil}

\author{S. P. Walborn}
\affiliation{Instituto de F\'{\i}sica, Universidade Federal do Rio de
Janeiro, Caixa Postal 68528, Rio de Janeiro, RJ 21941-972, Brazil}

\email[]{toscano@if.ufrj.br}

\begin{abstract}
A general procedure to construct criteria for identifying genuine multipartite continuous variable entanglement is presented. It relies on the definition of adequate global operators describing the multipartite system, the positive partial transpose criterion of separability, and quantum mechanical uncertainty relations. As a consequence, each criterion
encountered consists in a single inequality nicely computable and experimentally feasible. Violation of the inequality is sufficient condition for genuine multipartite entanglement. 
Additionally we show that 
the previous work of van Loock and Furusawa [Phys. Rev. A, {\bf 67}, 052315 
(2003)] is a special case of our result.
\end{abstract}




\maketitle

\section{Introduction}
\label{SectionIntro}
Genuine multipartite entanglement -- entanglement between three or more quantum systems -- is essential to harness the full power of quantum computing in either the circuit \cite{jozsa03} or one-way models \cite{raussendorf01} as well as to the security in multi-party quantum encryption protocols \cite{schauer10}. Additionally, it  provides increased precision in quantum metrology \cite{giovannetti04sci,Gross2010}, furnishes the resource to solve the Byzantine agreement problem 
\cite{neigovzen08}, and allows for multi-party quantum information protocols such as open destination quantum teleportation \cite{zhao04}.  There is also evidence that it is responsible for efficient transport in biological systems \cite{sarovar10}, and is linked to fundamental aspects of phase transitions  \cite{Sanpera2014}.  

Experimental identification of genuine multipartite entanglement is essential, since in order to reliably realize any multipartite entanglement-based task, it is necessary to confirm the presence of genuine multipartite entangled states.  Although quantum state tomography can provide all of the available information about the system, it requires a number of measurements that increases exponentially with the number of subsystems. 
Thus, entanglement witnesses composed of an abbreviated number of measurements are the most viable method for identifying genuine multipartite entanglement. This is true especially when the system is of high dimension, due to the dimension of the Hilbert space of each subsystem and/or the number of constituent subsystems.  
\par
In particular, 
for a continuous variable (CV) system composed of $n$ subsystems or modes, quantum state tomography is not viable in general.  This is rapidly becoming an important experimental concern, since possibly genuine multipartite CV entanglement  has been produced for three degenerate \cite{aoki03} and non-degenerate modes \cite{coelho09}, and more recently for a large number of {temporal \cite{yokoyama13} or spectral \cite{chen14,Gerke2015} modes.} Even for the specific case of two-modes, the
 difficulty of state tomography has also led to several criteria involving second-order \cite{simon00,duan00,mancini02,Giovannetti2003} and higher-order moments \cite{shchukin05,biswas05,hillery06a,walborn09,saboia11,Lee2014} of the canonical variables.  
 Most of these criteria are based on the positive partial transpose (PPT) argument \cite{peres96,horodecki96} 
and uncertainty relations involving the variance \cite{simon00,duan00,mancini02,Giovannetti2003} or entropy \cite{walborn09,saboia11} of marginal distributions. 
Motivated by the impossibility of PT based entanglement criteria to detect bound entanglement 
\cite{Horodecki2000}, in \cite{Sperling2009} the authors construct  bipartite
entanglement witnesses solving the so-called separability eigenvalue equation. The authors  
extend their approach in \cite{Sperling2013} deriving a set of algebraic equations,
which yield the construction of arbitrary multipartite
entanglement witnesses using general Hermitian operators. This approach is powerful when several
measurements of different Hermitian operators are considered in order to feed the optimisation problem 
involved to obtain the optimal entanglement criteria with the measurements at hand. Despite the lack of economy in terms of number of measurements, it was successfully implemented in \cite{Gerke2015}  to characterise multipartite entanglement in multimode frequency-comb Gaussian states, where the entire covariance matrix of the state was measured \cite{Roslund2013}.
\par
From a practical point of view, genuine entanglement criteria that are economical in terms of 
the number of measurement required for their implementation are more desirable.
A first step in this direction for CV systems was made by van Loock and Furusawa \cite{vanloock03} using a variance criterion to test full $n-$partite inseparability. 
 These criteria do not require reconstruction of the covariance matrix, and for this reason they were preferred to test  full $n-$partite inseparability in most of the CV multipartite states generated in the laboratory 
\cite{aoki03,Yukawa2008,coelho09,Pysher2011,Morizur2012,chen14}.
However, genuine multipartite entanglement is different from  $n-$partite inseparability, which can be 
produced by mixing quantum states with fewer that $n$ entangled sub-systems.
In general, to prove genuine multipartite entanglement, one must show that the state cannot be written as a convex sum of biseparable states  \cite{Bancal2011,reid13}.  
In reference \cite{reid13}, the authors show that one of the criterion of van Loock and Furusawa, 
consisting of a single inequality for a proper non-local operator, is indeed 
a genuine multipartite entanglement criterion, and extend the criterion also for
the product of variances. They also show how to adapt some of the criteria in \cite{vanloock03},
for three and four modes, in order to obtain genuine entanglement criteria.    
It is important to note that the criteria in \cite{vanloock03,reid13} are based on uncertainty relations (UR) only for the variances of non-local linear observables (NLO) ({\it i.e.} linear combination 
of canonical conjugate local observables).
\par
In this work, we solve the problem of how to construct genuine entanglement criteria for $n$ CV modes using the
{positive partial transposition (PPT) separability criterion in conjunction with general
uncertainty relations for non-local linear observables (URNLO).
We will call these  ``PPT+URNLO" criteria.
Our systematic method consists of adequate definitions of global operators of the $n$-partite system, which can be employed together  with a wide range of quantum mechanical uncertainty relations, producing classes of unique inequalities that test genuine $n$-partite entanglement.
In particular, we derive the unique family constituting single pairs of global $n$-mode position and momentum  operators that simultaneously test entanglement in all possible bipartitions and 
also genuine $n$-partite entanglement.
In the multipartite scenario, the criteria of van Loock and Furusawa \cite{vanloock03} and also the criteria in \cite{reid13,Shalm2012,Armstrong2015} constitute particular cases of our results. In the bipartite
scenario, we recover the most popular entanglement criteria \cite{simon00,duan00,mancini02,Giovannetti2003,walborn09,saboia11}.
\par 
The versatility of our results is twofold: i) for a fixed type of uncertainty relation we give the recipe 
of how to search all the pairs of non-local operators that could certify 
genuine multipartite entanglement  in a given prepared target state, ii) for a fixed pair or a set of pairs 
of non-local operators whose marginal distributions are available to the experimentalist, we provide the
possibility to search which uncertainty relation is more suitable to certify genuine multipartite entanglement.
\par
In addition to showing that the criteria in 
\cite{duan00,mancini02,Giovannetti2003,vanloock03,reid13,Shalm2012,Armstrong2015},
originally derived using the Cauchy-Schwarz inequality can be rederived using PPT separability criterion 
explicitly, we provide a general framework to obtain genuine entanglement criteria using PPT arguments 
and general uncertainty relations for non-local linear observables.
If we consider that most of the target states in quantum information tasks are pure and their experimental implementation are meant to approximate these states, mixed bound entangled states are practically excluded. 
Even for mixed states, bound entanglement  is a rare phenomena \cite{horodecki03}, so genuine 
entanglement criteria based on PPT arguments which are economical for their implementation are very useful.
\par 
This work is organised as follows. In Section \ref{sectionII} we establish the general idea of the class of
 ``PPT+URNLO" criteria and apply it to the case of two mode systems, and give several examples of well known bipartite entanglement criteria that belong to this class.  In section \ref{SectionIII} we set the notation to describe all the bipartitions of  an $n$-mode system and we review the definition of genuine multipartite  entanglement.  Section \ref{SectionIV} is devoted to the derivation of  ``PPT+URNLO"  entanglement criteria to test bipartite entanglement in the multipartite scenario.  In section \ref{SectionV} we use these bipartite criteria to build several criteria for genuine multipartite entanglement.  Concluding remarks are given in section \ref{SectionVI}.

}
\section{Bipartite Criteria for Two-mode Systems}
\label{sectionII}
Let's us begin introducing our approach in the case of a bipartite system.
For states $\hat \rho$ of two bosonic modes, several entanglement criteria have been developed to detect bipartite entanglement
\cite{simon00,duan00,mancini02,Giovannetti2003,Hyllus2006,walborn09,saboia11,agarwal05,hillery06a,shchukin05,Sperling2009,Lee2014}. Most of these criteria \cite{simon00,duan00,mancini02,Giovannetti2003,walborn09,saboia11,agarwal05,hillery06a,shchukin05,Lee2014} rely on the PPT separability criterion.  We note that several of these criteria \cite{duan00,mancini02,Giovannetti2003} were not originally derived using 
the PPT  separability criterion explicitly.  Nevertheless, in section \ref{sec:EBC} we show that they are indeed special cases of a general PPT separability criterion.  
\par
Some of these entanglement criteria  \cite{simon00,duan00,mancini02,Giovannetti2003,walborn09,saboia11,Lee2014} 
are constructed using quantum-mechanical uncertainty relations that are based on quantities associated with operators of the form 
\begin{subequations}
\label{eq:uv2}
\begin{align}
\hat u & =h_1\hat x_1+h_2\hat x_2 \\
\hat v & =g_1\hat p_1- g_2\hat p_2,
\end{align}
\end{subequations}
where $h_j$ and $g_j$ are arbitrary real numbers.  {We call these ``non-local" observables in the sense that they are linear combinations of observables on both systems.}  
In general, $\hat u$ and $\hat v$ do not commute: $[\hat u,\hat v]=i\gamma \hat {\bmath 1}$, where $\gamma = h_1g_1-h_2g_2$. Then,  a generic URNLO will be satisfied by all states $\hat{\rho}$, and can be written as
\beq
\label{UR}
F[\hat \rho,P_{\hat  {{u}}},P_{\hat {{v}}}]\geq f(|\gamma|),
\eeq
where $f$ is an increasing function of its argument and
$F$ is a functional, involving quantities that can be {either directly measured or}
determined from the marginal probabilities distributions $P_{\hat u}(u)=\bra{u} \hat \rho \ket{u}$ and $P_{\hat v}(v)=\bra{v} \hat \rho \ket{v}$ that are associated with the measurement of
$\hat u$ and $\hat v$ on the state $\hat \rho$.
\par
Using uncertainty relations of this type, a PPT+URNLO bipartite entanglement criteria can be derived in the following generic way.
A consequence of the PPT separability criterion \cite{peres96,horodecki96,simon00} is that 
if the partial transpose operator $\hat\rho^{T}$ \footnote{ Here, the operator $\hat\rho^{T}$ corresponds to
the partial transposition of the density matrix of $\hat \rho$, in some basis, with respect to one of the two modes.} does not correspond to 
a proper density operator ({\it i.e.} it has negative eigenvalues) then the original state is entangled with respect to that bipartition. This means that the URNLO
\beq
\label{criterion-general1}
F[\hat \rho^T,P_{\hat  {{u}}},P_{\hat {{v}}}]\geq  f(|\gamma|)
\eeq 
can be violated because $\hat u$ and $\hat v$ do not necessarily verify an uncertainty relation
when  the operator $\hat \rho^T$ does not correspond to a quantum state.  This only occurs when the original bipartite state  $\hat \rho$ is entangled. Violation of Eq.(\ref{criterion-general1}) is only a sufficient condition for bipartite entanglement, and is not a necessary one.  Nevertheless, criteria of this type are appealing due to the reduced number of measurements required, {and so they are
widely used in experimental detection of entanglement  in continuous  variable systems.}    
\par
{Partial transposition is a non-physical transformation. Thus,} inequality (\ref{criterion-general1}) is a useful entanglement criterion only if there is a simple way to obtain  $F[\hat \rho^T,P_{\hat  {{u}}}(\xi),P_{\hat {{v}}}(\xi)]$ from measurements on the original state $\hat \rho$.
{This connection can be made by considering the fact that transposition is equivalent to a mirror reflection in phase space \cite{simon00}, taking $(x,p) \longrightarrow (x,-p)$.  For two
mode states, non-local observables 
of the form 
\begin{subequations}
\label{eq:uv2}
\begin{align}
\hat \mu & =h_1\hat x_1+h_2\hat x_2 \\
\hat \nu & =g_1\hat p_1+g_2\hat p_2 
\end{align}
\end{subequations}
are then transformed as $\hat \mu \longrightarrow \hat u$ and $\hat \nu \longrightarrow 
\hat v=\pm g_1\hat p_1\mp g_2 \hat p_2$ where the plus (minus) sign corresponds to transposition on the second (first) mode.  
For this reason, throughout this paper we will refer to $\mu$, $\nu$ as the ``original" variables, and $u,v$ as the ``mirrored" variables. \par
With this correspondence between the original and mirrored variables, it can be shown that functionals related to probability distributions of $\mu$ and $\nu$ on the original state $\rho$ are equivalent to those related to $u$ and $v$ on the partially transposed state: 
\beq
\label{equality1}
F[\hat \rho^T,P_{\hat  {{u}}},P_{\hat {{v}}}]=F[\hat \rho,P_{\hat  {{\mu}}},P_{\hat {{\nu}}}].
\eeq 
Note that 
in general $\hat \mu$ and $\hat \nu$ also do not commute: $[\hat \mu,\hat \nu]=i\delta \hat {\bmath 1}$, where $\delta = h_1g_1+h_2g_2$.  Thus, they satisfy the equivalent  uncertainty relation 
\beq
\label{criterion-general2}
F[\hat \rho,P_{\hat  {{\mu}}},P_{\hat {{\nu}}}]\geq  f(|\delta|).
\eeq 
}
Combining Eqs.(\ref{criterion-general1}), (\ref{equality1}) and (\ref{criterion-general2}) we
can write any PPT+URNLO bipartite entanglement criterion as:
\beq
\label{criterion-general3}
F[\hat \rho^T,P_{\hat  {{u}}},P_{\hat {{v}}}]=F[\hat \rho,P_{\hat  {{\mu}}},P_{\hat {{\nu}}}]\geq  f(|\bar \delta|),
\eeq
where $|\bar \delta|=\max\{|\gamma|,|\delta|\}$.
Thus, for two mode systems we have $|\bar \delta|=|h_1g_1|+|h_2g_2|$.  
Due to the UR in Eq.(\ref{criterion-general2}) violation of the inequality 
in Eq.(\ref{criterion-general3}) is only possible in the cases when $|\bar \delta|=|\gamma|> |\delta|$ and implies entanglement.  Thus, it is necessary to choose $\gamma$ and $\delta$ adequately.

\subsection{Examples of Bipartite Criteria}
\label{sec:EBC}

Let us briefly review some PPT+URNLO bipartite entanglement criteria that appear in the literature.  In Table \ref{tab:1} we identify the correspondence between our notation and previous entanglement criteria.
\par
In the case where $F$ is the sum of variances $\Delta^2\hat \xi$ of the marginal distributions
$P_{\hat \xi}(\xi)$ ($\xi= \mu,\nu$),
we have the sum of variance entanglement criterion:
\beq
\label{linearUR}
F_{Lin}[\hat \rho,P_{\hat  {{\mu}}},P_{\hat {{\nu}}}]\equiv\Delta^2\hat \mu+\Delta^2\hat {\nu}\geq f_{Lin}(|\bar \delta|) = |\bar \delta|,
\eeq
that appears in Eq.(11) of Ref. \cite{simon00} and in Eq.(3) of Ref. \cite{duan00}
(the non-local observables  can be mapped in our notation through 
the identification showed in Table \ref{tab:1}).
\par
\begin{table*}[t]
\begin{tabular}{| l | l | l | l | l | l |}
\hline
Reference           & Equation   & Criterion & Order & ``Original'' non-local  
& ``Original'' non-local   
\\&&&&position:  $\hat \mu\equiv h_1\hat x_1+h_ 2 \hat x_2$     & momentum:
$\hat \nu\equiv g_1\hat p_1+g_2\hat p_2$               
\\ \hline
Simon \cite{simon00}    &  Eq.(11)  & $F_{Lin}$         & 2nd   & $h_1 \hat x_1=d_1\hat q_1+d_2 \hat p_1$ & $g_1 \hat p_1=d_1^{\prime}\hat q_1+d_2^{\prime} \hat p_1$
\\&&&&$h_2 \hat x_2=d_3\hat q_2+d_4 \hat p_2$&$g_2 \hat p_2=d_3^{\prime}\hat q_2+d_4^{\prime} \hat p_2$
\\ \hline
DGCZ \cite{duan00}      & Eq.(3)   & $F_{Lin}$         & 2nd   & $ h_1=|a|,h_2=\frac{1}{a}$   & $h_1=|a|,g_2= - \frac{1}{a}$, with real $a$ 
\\ \hline
MGVT \cite{mancini02}   & Eq.(6)    & $F_{H}$           & 2nd   & $h_1=1,h_2=1$               & $g_1=1,g_2=-1$               \\ \hline
GMVT \cite{Giovannetti2003}  & Eq.(28) & $F_{H}$           & 2nd   & $h_1 \hat x_1=a_1\hat q_1+a_3 \hat p_1$ &  $g_1 \hat p_1=b_1\hat p_1+b_3 \hat q_1$
\\&&&& $h_2 \hat x_2=a_2\hat q_2+a_4 \hat p_2$&
 $g_2 \hat p_2=b_2 \hat p_2+b_4 \hat q_2$
\\ \hline
WTSTM \cite{walborn09}    & Eq.(13) & $F_{E}$           & $>2$  & $h_1=1,h_2=\pm 1$               & $g_1=1,g_2=\mp 1$               
\\ \hline
STW \cite{saboia11}      &  Eq.(16) & $F_{E}$           & $>2$  & $h_1=1,h_2=\pm 1$               & $g_1=1,g_2=\mp 1$               \\ \hline
\end{tabular}
\caption{\label{tab:1} Popular bipartite entanglement criteria that are  of the type we call PPT+URNLO. 
In all cases the mirrored operators are (see the text): the non-local position 
$\hat u=\hat \mu $ and the non-local momentums $\hat v=-g_1\hat p_1+g_2\hat p_2$ (if the partial transposition is with respect of the first mode) or 
$\hat v=g_1\hat p_1-g_2\hat p_2$  (if the partial transposition is with respect of the second mode). 
We indicate the order of the moments of the ``original'' operators that appear in each criterion.
Note that tabulated operators on the l.h.s. are ours.}
\end{table*}
\par
The product of variance entanglement criterion given in Eq.(6) of Ref. \cite{mancini02} and in Eq.(28) of Ref. \cite{Giovannetti2003} can be obtained 
if we choose the functional $F$ as:
\beq
\label{HeisenbergUR}
F_{H}[\hat \rho,P_{\hat  {{\mu}}},P_{\hat {{\nu}}}]
\equiv \Delta^2 \hat \mu \Delta^2 \hat \nu \geq
f_H(|\bar \delta|)= 1/4|\bar\delta|^2.
\eeq
(see Table \ref{tab:1} for the identification of the non-local observables 
in references \cite{mancini02,Giovannetti2003} within our notation).
The Shannon-entropic entanglement criterion is obtained when the functional $F$ is
the sum of Shannon entropies: 
\bea
F_{E}[\hat \rho,P_{\hat  {{\mu}}},P_{\hat {{\nu}}}]&=& h[P_{\hat \mu}]+ h[P_{\hat {\nu}}]\geq\nonumber\\
&\geq& f_E(|\bar \delta |) =  \ln (\pi e|\bar \delta|),
\label{entropicUR}
\eea
$h[P] \equiv -\int dx \;P(x) \; \ln(P(x))$ is the Shannon entropy of the probability distribution function $P(x)$.
This new criterion presented here is based on a new uncertainty relation, 
$h[P_{\hat u}]+ h[P_{\hat v}]\geq \ln (\pi e|[\hat u,\hat v]|)$, recently proved in Ref. \cite{Huang2011} for generic 
operators $\hat u$ and $\hat v$ in an $n$-mode bosonic system that are generic linear combination of the position and momentum of each mode.
In the particular case when the mirrored operators $\hat u, \hat v$ form conjugate pairs, 
{\it i.e.} they are related by a $\pi/2$ rotation therefore the wavefunctions corresponding to the eigenstates $\ket{u}$ and $\ket{v}$ are related by a Fourier Transform, we recover the bipartite entanglement
criteria  in Eq.(13) of \cite{walborn09} that used the old entropic uncertainty relation  for conjugate pairs of operators derived in \cite{bialynicki75} (see Table \ref{tab:1} for the identification of the non-local operators within our notation).
Nevertheless, here we
have extended the result in \cite{walborn09} because  the Shannon-entropic entanglement criterion given  in  our Eq.(\ref{entropicUR}) is valid for generic linear non-local operators $\hat \mu$ and $\hat \nu$
not necessarily conjugate pairs.
\par
The three entanglement criterion given in Eqs.(\ref{linearUR}), (\ref{HeisenbergUR})
and (\ref{entropicUR}) can be written all together in a single inequality:
\bea
\label{all-inequalities}
&&\ln\left[\pi e (\Delta^2 \hat \mu+\Delta^2\hat \nu)\right]\geq
\ln\left(2\pi e \Delta \hat \mu \Delta\hat \nu \right) \geq \nonumber\\
&&\geq h[P_{\hat \mu}]+ h[P_{\hat \nu}]  \geq \ln (\pi e |\bar \delta|).
\eea
In this way, we see that the Shannon-entropic entanglement criterion is the strongest criterion because its corresponding inequality can be  violated in cases when the other two are not, thus it can detect bipartite entanglement when the other two do not. Exemples of bipartite states
whose entanglement can be detected with the Shannon-entropic entanglement criterion but 
can't be detected with either variance criteria
were shown in Ref. \cite{walborn09}. 
\par
The examples presented here show that in order to create PPT+URNLO bipartite entanglement criteria we only need URs of the form Eq. (\ref{UR}), valid for any pair  $\hat u$ and $\hat v$ of non-commuting linear non-local operators .  
However, the most common UR relations are defined for a single bosonic mode. Nevertheless, it is straightforward to use this type of UR if we restrict ourselves to conjugate pairs
$\hat u$ and $\hat v$ of non-local operators because these operators define a type of 
``non-local'' mode of its own right. 
For example, the UR in terms of the Rényi entropies were used in Eq.(16) of \cite{saboia11} to create a  PPT+URNLO bipartite entanglement criterion. We recall  the UR in terms of the Rényi entropies:
$h_{\alpha}[P]= \frac{1}{1-\alpha}\ln \left[\int {\rm d}x P^\alpha (x) \right ]$, 
derived in \cite{bialynicki06}:
\beq
h_{\alpha}[P_{\hat \mu}]+ h_{\beta}[P_{\hat {\nu}}] \geq
    -\frac{1}{2(1-\alpha)} \ln \frac{\alpha}{\pi|\gamma|}-\frac{1}{2(1-\beta)} \ln \frac{\beta}{\pi|\gamma|},
    \label{eqBirulapm}
\eeq
that is valid for any conjugate pairs with $[\hat u,\hat v]=i\gamma \hat {\bmath 1}$ and $1/\alpha+1/\beta=2$.
{\color{green}
}

\section{Definition of genuine multipartite entanglement}
\label{SectionIII}
{The same concepts outlined in the last section can be used to develop entanglement criteria for multipartite systems.   A multipartite system consisting of $n$ parts, can be entangled in a number of different ways. For example, an $n$-partite state is
said to be \emph{genuinely} $n$-partite entangled if it cannot be prepared by mixing states
that are separable with respect to some bipartition, dividing the constituent sub-systems into two groups.}   
Thus, in order to define genuine multipartite entanglement in an  $n$-mode state $\hat \rho\in \otimes_{i=1}^{n} {\cal H }_i$, we 
need to fix the notation to specify the different possible bipartitions of the 
system.   
We denote 
 a bipartition of the system by $\vec \alpha | \vec \beta$, where the vectors of integer indexes 
$\vec \alpha\equiv(\alpha_1,\ldots,\alpha_{n_A})$
and $\vec \beta\equiv(\beta_1,\ldots,\beta_{n_B})$ indicate the modes belonging to each part, 
and $ \alpha_i,\beta_i$ are integers in the set $\{1,\ldots,n\}$.
Thus, part $A$ of the bipartition denoted by $\vec \alpha$, has $n_A$ modes and part $B$ denoted by $\vec \beta$ has $n_B=n-n_A$ modes.
For convenience we order $\alpha_{i}<\alpha_{i+1}$ ($i=1,\ldots,n_A$) and $\beta_{j}< \beta_{j+1}$ 
($j=1,\ldots,n_B$).
\par
The different bipartitions of an $n$ mode system can be classified according to
the number of modes in each
part.  Thus, the class $(n_A,n_B)$ contains all possible bipartitions with $n_A$ modes in  part $A$
and $n_B$ modes in part $B$. 
We call $n_A^{\mbox{\scriptsize max}}$ the number of different bipartition classes,
{\it i.e.} $n_A=1,\ldots,n_A^{\mbox{\scriptsize max}}$ where $n_A^{\mbox{\scriptsize max}}=n/2$
if $n$ is even and $n_A^{\mbox{\scriptsize max}}=(n-1)/2$ if $n$ is odd.
In a given class $(n_A,n_B)$ there are $N_{n_A}$ bipartitions that correspond to different 
labels $\vec \alpha | \vec \beta$.  For $n_A \neq n/2$,
$N_{n_A}= {n\choose n_A}$, and for $n_A=n/2$, $N_{n_A}=\frac{1}{2} {n\choose n_A}$.
It is easy to see that for either $n$ odd or even, there are a total of $L=2^{n-1}-1$ different bipartitions.
For example, in the case of a four mode system ($n=4$), we have two classes ($n_A^{\mbox{\scriptsize max}}=2$): i) $(n_A=1,n_B=3)$ that corresponds to  the
$N_{n_B}=4$ bipartitions,  $\vec \alpha |\vec \beta=1|234,2|134,3|124,4|123$,
and ii) $(n_A=2,n_B=2)$ that corresponds to the $N_{n_B}=3$ bipartitions,
$\vec \alpha |\vec \beta=12|34,13|24,14|23$.
The total number of bipartitions in this case is $L=N_{n_A=1}+N_{n_A=2}=4+3=2^{4-1}-1=7$.
The table \ref{tab:2} presents some examples with $n=1,\ldots,6$ for the number of classes ($n_A^{max}$) and partitions ($L$).

\begin{table}
\begin{tabular}{|c|c|c|}
\hline
$n$ (modes) & $n_A^{max}$ (classes) & $L$ (partitions)\\ \hline
2           & 1                     & 1               \\ \hline
3           & 1                     & 3               \\ \hline
4           & 2                     & 7               \\ \hline
5           & 2                     & 15              \\ \hline
6           & 3                     & 31              \\ \hline
\vdots      & \vdots                & \vdots          \\ \hline
$n$         & $n/2$ for $n$ even, $(n-1)/2$ for $n$ odd & $L=2^{n-1}-1$    \\ \hline
\end{tabular}
\caption{\label{tab:2} Number of modes, classes, and partitions.}
\end{table}

\par
It was a great advance in the understanding of multipartite entanglement to distinguish 
full $n$-partite inseparable states {(definition below)} from those states that have genuine multipartite entanglement
\cite{Bancal2011,Shalm2012,reid13}.
A genuine $n$-partite entangled state is one that does not belong to the family
of ``biseparable'' states \cite{Bancal2011,reid13}:
\bea
\hat \rho_{bs}&\equiv &\sum_{\{\vec \alpha |\vec \beta\}}
p_{\{\vec \alpha|\vec \beta \}}\hat{\rho}_{\{\vec \alpha|\vec \beta\}}=\nonumber\\
&=&\sum_{\{\vec \alpha|\vec \beta \}}
p_{\{\vec \alpha|\vec \beta \}}\left(\sum_{j}\eta_j^{\{\vec \alpha|\vec \beta\}}\hat \rho^{\{\vec \alpha\}}_j\otimes\hat \rho^{\{\vec \beta\}}_j\right),
\label{biseparable-states}
\eea
where the first sum in Eq.(\ref{biseparable-states}) runs over the set $\{\vec \alpha|\vec \beta\}$ of all the $L$ possible bipartition's of the system and  the states in between parenthesis are generic separable states in the bipartition $\vec \alpha|\vec \beta$.   Normalization requires that  $\sum_{\{\vec \alpha |\vec \beta\}}
p_{\{\vec \alpha|\vec \beta \}}=1=\sum_j \eta_j^{\{\vec \alpha|\vec \beta\}}$.  {Note that for a given class, all the modes in a given part might be entangled. Thus, all forms of genuine $n_A$- or $n_B$-partite entanglement may appear in state \eqref{biseparable-states}.
}
\par
Any genuine multipartite entanglement criterion needs to test entanglement in all the possible bipartitions that can be drawn in the system. However, testing bipartitions individually is not enough, since 
biseparable states \eqref{biseparable-states} could be entangled in every bipartition of the system. These are called full $n$-partite inseparable states \cite{Shalm2012,reid13}.  For example, a class of  $n$-partite inseparable states corresponds to biseparable states with all the coefficients
$p_{\{\vec \alpha|\vec \beta \}}$ different from zero in Eq.(\ref{biseparable-states}),
although this is not a necessary condition (see a $3$-mode example in \cite{Shalm2012}).
Therefore, a genuine entanglement criteria needs to refute biseparable state 
$\hat \rho_{bs}$ as a possible description of the system.
\par
{In the following section we first introduce entanglement criteria to test entanglement in any bipartition of an $n$-mode system, and then use these criteria to construct criteria for genuine $n$-partite entanglement. }

\section{Entanglement criteria for Generic Biparitions}
\label{SectionIV}
{Here we consider that the multipartite system consists of $n$ bosonic modes, described through local canonical 
operators:
 \beq
 \vectg{\hat z}\equiv(\bsy {\hat x},\bsy {\hat p})^T=(\hat x_1,...,\hat x_n,\hat p_1,...,\hat p_n)^T,
 \label{eq:z}
 \eeq
  where $T$ means transposition.
 Each pair of canonically conjugate observables with continuous spectra 
 {$ x_j$ and $p_j$}  verifies the commutation relation  $[\hat x_j,\hat p_k]=i \delta_{jk}$ and we generically call them  position and momentum, respectively.}
Here we'll prove that  the general form introduced in Eq.(\ref{criterion-general3}) for bipartite entanglement criteria can be applied not only for a two mode system
but also for the general case of an arbitrary bipartition $\vec \alpha|\vec \beta$ of an $n$-mode bosonic system.  We indicate the partial transposition as $T_{\bfal}$, with
$\bfal=\vec \alpha$ if the partial transposition is with respect to the modes contained in $\vec \alpha$
and $\bfal=\vec \beta$ if the partial transposition is with respect to the modes contained in $\vec \beta$. 
Because each of the $L=2^{n-1}-1$ possible bipartitions can be tested with a different pair of non-local operators, we rewrite Eq.(\ref{criterion-general3}) as
\beq
F[\hat \rho^{T_{\bfal}},P_{\hat u_{\bfal}},P_{\hat v_{\bfal}}]=
F[\hat\rho ,P_{\hat \mu_{\bfal}},P_{\hat \nu_{\bfal}}]\geq  f(|\bar \delta_{\bfal}|),
\label{criterion-general3-rewrite}
\eeq
where $|\bar \delta_{\bfal}|=\max\{|\gamma_{\bfal}|,|\delta_{\bfal}|\}$ with $[\hat u_{\bfal},\hat v_{\bfal}]=i\gamma_{\bfal}\hat{\bmath 1}$ and 
$[\hat \mu_{\bfal},\hat \nu_{\bfal}]=i\delta_{\bfal}\hat{\bmath1}$.
The objective here is to determine what entanglement criteria can be tested, given that the experimentalist has access to marginal probability distributions of observables \eqref{new-operators}. 
Therefore, in the following we will show that for a fixed pair of operators:
\bse
\label{new-operators}
\bea
\label{new-position-type}
\hat {{\mu}}_{\bfal}&\equiv&\sum_{j=1}^n h_j\hat x_j,\\
 \label{new-momentum-type}
\hat{ {\nu}}_{\bfal}&\equiv&
\sum_{j=1}^n  g_j\hat p_j,
\eea
\ese
with commutator 
\beq
\label{delta-bfal}
\delta_{\bfal}=\sum_{j=1}^n h_jg_j,
\eeq
where $h_j, g_j$  are  arbitrarily real numbers, it is always possible to find mirrored operators $\hat u_{\bfal}$ and $\hat v_{\bfal}$ that satisfy the equality in Eq. \eqref{criterion-general3-rewrite}.
But, first note that if we want to accomodate observables of the form $\hat {{\mu}}_{\bfal}=\sum_{j=1}^n h^{\prime}_j\hat x^{\prime}_j+g^{\prime}_j\hat p^{\prime}_j$ and 
$\hat {{\nu}}_{\bfal}=\sum_{j=1}^n h^{\prime\prime}_j\hat x^{\prime}_j+g^{\prime\prime}_j\hat p^{\prime}_j$, we simply redifine the quadrature variables as $h_j\hat x_j=h^{\prime}_j\hat x^{\prime}_j+g^{\prime}_j\hat p^{\prime}_j$
and $g_j\hat p_j=h^{\prime\prime}_j\hat x^{\prime}_j+g^{\prime\prime}_j\hat p^{\prime}_j$.
\par

All of the possible non-local operators of the $n$-mode system are defined as linear combinations of the local operators written in \eqref{eq:z}.   We can write them as
\beq
\label{auxiliary-old-operators}
\vectg{\hat { z}}_{\bfal} \equiv (\vectg{\hat u}_{\bfal},\vectg{\hat {v}}_{\bfal})^T
=\bmath{M}_{\bfal}\vectg{\hat z},
\eeq
where 
\beq
\label{auxiliary-old-operators}
(\vectg{\hat u}_{\bfal},\vectg{\hat {v}}_{\bfal})^T\equiv
(\hat u_{1,\bfal},...,\hat u_{n,\bfal},\hat {v}_{1,\bfal},...,\hat {v}_{n,\bfal})^T,
\eeq
is the $2n$-component vector of possible linear combinations.
The matrix $\bmath{M}_{\bfal}\equiv\ diag(\Mxal,\Mpal)$ is a $2n \times 2n$ real matrix and
$\Mxal$ and $\Mpal$ are non-singular real $n \times n$ matrices that we call the x-matrix 
and the p-matrix of the bipartition, respectively.  
First, we identify the mirrored observables in  Eq.(\ref{criterion-general3-rewrite}) as
\bse
 \label{old-operators}
 \bea
 \hat u_{\bfal}&\equiv&\hat u_{k,\bfal}=\sum_{j=1}^n(\Mxal)_{k,j}\hat x_j,\\
 \hat v_{\bfal}&\equiv&\hat v_{k,\bfal}=\sum_{j=1}^n(\Mpal)_{k,j}\hat p_j.
 \eea
 \ese 
What remains is to determine the rows of the matrix $\mathbb{M}_{\bfal}$ such that the equality 
 $F[\hat \rho^{T_{\bfal}},P_{\hat u_{\bfal}},P_{\hat v_{\bfal}}]=F[\hat \rho,P_{\hat \mu_{\bfal}},P_{\hat \nu_{\bfal}}]$ in Eq.(\ref{criterion-general3-rewrite}) holds for any functional and for the measureable operators \eqref{new-operators}.
\par
Since the local operators satisfy $[\hat{x}_j,\hat{p}_k]=i \delta_{jk}$, we impose that the non-local operators satisfy the commutation relation 
$[\hat u_{j,\bfal},\hat {v}_{k,\bfal}]=i \gamma_{\bfal} \delta_{jk}$ with $\gamma_{\bfal}$ any real number. This means that: $\Mpal^T\Mxal=\Mxal \Mpal^T=\gamma_{\bfal} \bmath 1$ 
\footnote{Please note that this is not a symplectic condition over the matrix $\bmath{M}_{\bfal}$
that would corresponds to the condition $\Mpal^T\Mxal=\Mxal \Mpal^T=\bmath 1$.},
 so
the x-matrix is determined by the p-matrix:
\beq
\label{matrixMbfal}
\Mxal=\gamma_{\bfal} (\Mpal^{-1})^T.
\eeq
When the matrix $\bmath{M}_{\bfal}$ is orthogonal, {\it i.e.} 
$\bmath{M}_{\bfal}^T=\bmath{M}_{\bfal}^{-1}$,
we have that each pair of 
non-local observables $\hat u_{j,\bfal}$ and $\hat {v}_{j,\bfal}$ ($j=1,\ldots,n$) are conjugate operators.  In this particular case Eq.(\ref{matrixMbfal}) reduce to:
\beq
\label{matrixMbfal-orthogonal}
\Mxal=\gamma_{\bfal} (\Mpal).
\eeq

 In Appendixes \ref{Appendix1} and \ref{Appendix2}, we prove that the coefficients in Eqs. \eqref{old-operators} are uniquely given by the $k^{\mathrm{th}}$ row: 
 \beq
 \label{Mp-k-row-text}
 (\Mpal \bmath)_{kj}=\bar g_j=\left\{
    \begin{matrix} 
       -g_j & \mbox{if $j$ is one component of $\bfal$ }  \\
       g_j & \mbox{otherwise }\\
    \end{matrix}
 \right.
 \eeq 
 and 
\beq
 ((\Mpal^{-1})^T)_{k,j} = \frac{h_j}{\gamma_{\bfal}}. 
  \label{hj-text}\\
 \eeq 
 Therefore, the commutator between the mirrored observables is $[\hat u_{\bfal},\hat v_{\bfal}]=i\gamma_{\bfal}\hat{\bmath 1}$, with
 \beq
 \label{gamma-bfal}
 \gamma_{\bfal}=\sum_{j=1}^{n}h_j\bar g_j=  
\pm\gamma_{\vec \alpha},
 \eeq
where the plus sign is when $\bfal=\vec \alpha$ and the minus sign is when $\bfal=\vec \beta$. 
 \par
In this way,  we see that the coefficients of the mirrored  operators $\hat u_{\bfal}$ and $\hat v_{\bfal}$ are determined once we specify the p-matrix of the bipartition $\Mpal$, whose $k^{\mathrm{th}}$ row
is given in Eq.(\ref{Mp-k-row-text}) and with the  $k^{\mathrm{th}}$ row of $(\Mpal^{-1})^T$ given
in Eq.(\ref{hj-text}).
Without loss of generality we can choose $k=1$. In 
Appendix \ref{Appendix2}  we give the general structure of a matrix $\Mpal$ with these properties that satisfied \eqref{matrixMbfal}. 
This proves the equality 
$F[\hat \rho^{T_{\bfal}},P_{\hat u_{\bfal}},P_{\hat v_{\bfal}}]=F[\hat \rho,P_{\hat \mu_{\bfal}},P_{\hat \nu_{\bfal}}]$ in Eq.(\ref{criterion-general3-rewrite}).
\par
For states $\hat \rho_{\{\vec \alpha |\vec \beta\}}$ that are separable in the bipartition $\vec \alpha|\vec \beta$, the partial transposed state $\hat \rho_{\{\vec \alpha |\vec \beta\}}^{T_{\bfal}}$ (with $\bfal=\vec \alpha$ or $\bfal=\vec \beta$) is also a physical state.  In this case, 
Eq.(\ref{criterion-general3-rewrite}) is not violated, since it is a valid uncertainty relation for both sets of operators.
Therefore, violation of Eq. (\ref{criterion-general3-rewrite}) constitutes a bipartite entanglement 
criterion for $n$ modes in bipartition $\vec \alpha|\vec \beta$, where the lower bound is given by
\beq
|\bar \delta_{\bfal}|=\max\{|\gamma_{\bfal}|,|\delta_{\bfal}|\}=
\sum_{j\notin \{\bfal\}}|h_jg_j|+\sum_{j\in \{\bfal\}}|h_j\bar g_j|. 
\eeq
Here the first sum runs over indexes $j$ corresponding to those modes that were not transposed and the second over those modes that were transposed.
To observe the violation of inequality (\ref{criterion-general3-rewrite})  for states entangled in the bipartition it is necessary to choose the coefficients of the operators ($h_j$ and $g_j$) in a way such that 
 $|\bar \delta_{\bfal}|=|\gamma_{\bfal}|> |\delta_{\bfal}|$.
 \par 
 When we apply the sum of variance functional 
 $F=F_{Lin}$ defined in Eq.(\ref{linearUR}) with $f(|\bar \delta_{\bfal}|)=|\bar \delta_{\bfal}|$ in Eq.(\ref{criterion-general3-rewrite})
 we recover the  van Loock and Furusawa entanglement criterion in Eq. (28) of Ref. \cite{vanloock03}.
 The factor of $1/2$ of the lower bound  in  Ref. \cite{vanloock03} is due to the fact that there the commutation relation is $[\hat x_j,\hat p_k]=i\delta_{jk}/2$ instead of $[\hat x_j,\hat p_k]=i\delta_{jk}$, as defined here.
 We stress that although  the PPT argument is not used explicitely in Ref. \cite{vanloock03}, here we have shown these sum of variance inequalities indeed belong to the class of PPT+URNLO bipartite entanglement criteria. 
 \par
 We stress that we have proved that for every pair of  linear non-local observables 
 $(\hat \mu_{\bfal}, \hat \nu_{\bfal})$, it is always possible to find  a pair of  mirrored linear non-local observables $(\hat u_{\bfal}, \hat v_{\bfal})$ such that the equality in Eq. \eqref{criterion-general3-rewrite} holds.
This relies on the structure of the matrix $\Mpal$ in Appendix \ref{Appendix2}. 
In particular, we can always choose to test entanglement with  any pair of commuting  non-local observables
 ({\it i.e.} $[ \hat \mu_{\bfal}, \hat \nu_{\bfal}]=\delta_{\bfal}=0$), with the 
 advantage that in this case we simply have a lower bound given by $|\bar \delta_{\bfal}|=|\gamma_{\vec \alpha}|$. Moreover, the inequality can always be violated by a simultaneous eigenstate of the commuting operators, which are entangled states, and are the backbone of quantum information in CV systems  \cite{VanLoock2000,braunstein05}. Examples of these type of  eigenstates 
are the $2$-mode EPR states that are simultaneous  eigenstates
of relative position $\hat u=\hat x_1-\hat x_2$ and the total momentum 
$\hat v=\hat p_1+\hat p_2$ \cite{epr35}, or the CV GHZ $n$-modes states that are 
simultaneous  eigenstates of the relative positions 
$\hat u_1=\hat x_1-\hat x_2,\hat u_2=\hat x_2-\hat x_3,\ldots,\hat u_{n-1}=\hat x_{n-1}-\hat x_n$ and the total momentum $\hat v=\hat p_1+\ldots+\hat p_n$
\cite{vanloock03}. It is worth noting that these type of unnormalised eigenstates are well 
approximated by squeezed multipartite gaussian states that have been generated in several experiments recently \cite{yokoyama13,chen14,Gerke2015}. 
In the next section we use this type of commuting non-local linear operators to derive  criteria that are useful to detect genuine multipartite entanglement in any
$n$-partite state.

\section{Genuine multipartite PPT+URNLO entanglement criteria}
\label{SectionV}

\par
Here we will derive genuine PPT+URNLO entanglement criteria that exclude the possibility of a 
violation by the set of biseparable states $\hat \rho_{bs}$ given in Eq.(\ref{biseparable-states}). 
We will provide two different types of genuine entanglement criteria: 
a first one based on a single pair of commuting operators and a second one
based on a set of pairs of commuting operators.

\subsection{Criteria with a single pair  of commuting non-local operators}

Let us suppose that we can find a pair of suitable  non-local operators $\hat \mu=\sum_{j=1}^n h_j \hat x_j$ and $\hat \nu=\sum_{j=1}^n g_j \hat p_j$ such that
$[\hat \mu,\hat \nu]=0$ with
$\gamma_{\vec \alpha}=\sum_{j=1}^n h_j \bar g_j\neq 0$ for all the $L$ bipartition's of the system
(where  $\bar g_j=-g_{j}$ if  $j$ is one component of the vector $\vec \alpha$ or  $\bar g_j=g_j$ otherwise).
This means we must  consider all
possible different location of minus sign in $\bar g_j$  within $\gamma_{\vec \alpha}$ 
defined in Eq.(\ref{gamma-bfal}), 
for fixed values 
of the coefficients $h_j$ and $g_j$. 
Then in this case we could test entanglement in all the bipartition of the system through the different inequalities (see Eq.(\ref{criterion-general3-rewrite})),
\beq
\label{genuine-same}
F[ \hat \rho,P_{\hat \mu},P_{\hat \nu}]\geq f(|\gamma_{\vec \alpha}|),
\eeq 
but with a single pair $(\hat \mu,\hat \nu)$ of non-local operators. 
Before we prove that there is always a family of such pairs in an $n-$mode system, let us see that if $\gamma_{\mbox{\scriptsize min}}=\mbox{ min}_{\{\vec \alpha\}} \left\{|\gamma_{\vec \alpha}|\right\}\geq 0$, and $\{\vec \alpha\}$ runs over all the $L$ bipartitions of the system, then the single inequality,
\beq
\label{genuine-criteria}
F[\hat \rho,P_{\hat \mu}, P_{\hat \nu}]\geq
f(\gamma_{\mbox{\scriptsize min}})\geq 0,
\eeq
is never violated by biseparable states defined in Eq. (\ref{biseparable-states}).
So, violation 
of this single inequality constitutes a genuine multipartite entanglement criterion because it tests entanglement in all the bipartitions of the system and it is never violated by biseparable states. 
In order to prove this, one of  two conditions must be true:
i) the functional $F$ is concave with respect to 
a convex sum of density operators, {\it i.e.} 
$F\left[\hat \rho=\sum_j p_j \hat \rho_j,P_{\hat \mu},P_{\hat \nu}\right]\geq \sum_j p_j 
F[ \hat \rho_j,P_{\hat \mu},P_{\hat \nu}]$
or ii) the functional $F\left[\hat \rho,P_{\hat \mu},P_{\hat \nu}\right] \geq 
F^{\prime}\left[\hat \rho,P_{\hat \mu},P_{\hat \nu}\right]$ where $F^{\prime}$ is concave with respect to a convex sum of density operators.
If the first condition is valid, then for biseparable states $\hat \rho_{bs}$  we have,
\bea
F\left[\hat \rho_{bs},P_{\hat \mu},P_{\hat \nu} \right] &\geq & \sum_{\{\vec \alpha|\vec \beta \}}
p_{\{\vec \alpha|\vec \beta \}}F\left[\hat \rho_{\{\vec \alpha|\vec \beta \}},P_{\hat \mu},P_{\hat v}\right]\nonumber\\
&\geq&\sum_{\{\vec \alpha|\vec \beta \}}
p_{\{\vec \alpha|\vec \beta \}}f(|\gamma_{\vec \alpha}|)\nonumber \\
& \geq &
f(\gamma_{\mbox{\scriptsize min}})  \geq 0,
\label{demostration-no-violation}
\eea
where we assume by hypothesis that the inequality in Eq.(\ref{genuine-same}) is never violated by separable states $\hat \rho_{\{\vec \alpha|\vec \beta \}}$ in the bipartition $\vec \alpha|\vec \beta$.
Therefore, it can be seen immediately that the same is true if the second condition is valid. 
In Appendix \ref{Appendix3} 
we prove the concavity of the sum of entropies functional $F_E$.
This result, together with the chain of inequalities in Eq.(\ref{all-inequalities}), proves that, besides the entropic UR, we can use either the sum of variances UR in 
Eq.(\ref{linearUR}) or  the product of variances UR in Eq.(\ref{HeisenbergUR}) to set the functional $F$ in our 
genuine entanglement criterion in Eq.(\ref{genuine-criteria}).
\par 
In what follows we develop a systematic way to find  the non-local commuting operators in the genuine entanglement criterion in Eq.(\ref{genuine-criteria}). 
Let's start by defining $\bmath \Lambda_{\bfal}$ 
as the diagonal matrix with ones in the location of modes that
are not transposed and negative ones in the location of modes that are transposed.  We consider a ``seed" partition, defined by 
the vector $\bfbe \equiv(1,\ldots,n_A)$ of modes in $A$.
From this seed partition corresponding to a given class $(n_A,n_B)$, we can generate all elements of this class by swapping the modes around.  To do this we define the bipartition permutation matrix  
$\bmath{P}_{\bfbe\vec \alpha }=\prod_{i=1}^{n_A}\bmath P_{i\,\alpha_i}$,
where $\bmath P_{i\,\alpha_i}$ is the permutation matrix between mode $i$ and mode $\alpha_i$.  In other words, $\bmath P_{i\,\alpha_i}$ is   
the matrix obtained from swapping the rows $i$ and $\alpha_i$ of the identity matrix $\bmath{1}$,  and
$\bmath P_{i,i}=\bmath{1}$. 
Note that $\bmath{P}_{\bfbe\vec \alpha }\bmath{P}_{\bfbe\vec \alpha }^T=
\bmath 1$ because $\bmath P_{i\,\alpha_i}^2=\bmath 1$.
Therefore, for an arbitrary matrix $\bmath M$, the matrix $\bmath{P}_{\bfbe\vec \alpha }\bmath M$
corresponds to swapping all the rows in $\bmath M$ indicated by $\bfbe$ with all the target rows indicated
by $\vec \alpha$. Analogously, the matrix $\bmath M \bmath{P}_{\bfbe\vec \alpha}^T$
corresponds to swapping all the columns  in $\bmath M$ indicated by $\bfbe$ with all the target columns indicated
by $\vec \alpha$. 
\par
First, we observe that we can write the matrix $\bmath \Lambda_{\bfal}$,
associated with the partial transposition with respect to any set of modes indicated in the vector $\bfal$, as:
\beq
\bmath \Lambda_{\bfal}=\pm\bmath{P}_{\bfbe\vec \alpha }\bmath \Lambda_{\bfbe}
\bmath{P}_{\bfbe\vec \alpha }^T,
\label{gene-form-lambda}
\eeq
where the minus sign applies when $\bfal=\vec \beta$ and the plus sign
when $\bfal=\vec \alpha$. 
\par
We call $\bmath \Lambda_{\bfbe}$, $\bmath M_{p,\vec s}$ and
$\Mxbe=\gamma_{\bfbe} (\Mpbe^{-1})^T$ the ``seed'' matrices associated with the bipartition class $(n_A,n_B)$.
With these matrices we can express the operators defined in Eqs. \eqref{new-operators} for the seed bipartition as
\beq
\label{new-seed-operators}
\hat {{\mu}}_{1,\bfbe}=\sum_{j=1}^{n}\;(\Mxbe )_{1j} \hat x_j \;\;,\;\;
\hat {{\nu}}_{1,\bfbe}=\sum_{j=1}^{n}\;(\Mpbe \bmath \Lambda_{\bfbe})_{1j} \hat p_j.
\eeq
Furthermore, we impose that $[\hat {{\mu}}_{1,\bfbe},\hat {{\nu}}_{1,\bfbe}]=i(\Mpbe \bmath \Lambda_{\bfbe} \Mxbe^T)_{11}=i\delta_{\bfbe}=0$.
Therefore, according to Eqs.(\ref{old-operators}),  the mirrored non-local operators  are:
\beq
\label{old-operators-seed}
\hat { u}_{1,\bfbe}=\hat {{\mu}}_{1,\bfbe}\;\;\mbox{and}\;\;
\hat { v}_{1,\bfbe}=\sum_{j=1}^{n}\;(\Mpbe)_{1j} \hat p_j,
\eeq
such that $[\hat { u}_{1,\bfbe},\hat { v}_{1,\bfbe}]=i(\Mpbe  \Mxbe^T)_{11}=
i\gamma_{\bfbe}\neq 0$. 
We can rephrase these statements by saying that given the coefficient of the mirrored  
operators $\hat u_{\bfbe}$ and $\hat v_{\bfbe}$ corresponding to the first columns of the seed matrices  $(\Mxbe)_{1j}=(h_1,\ldots,h_{n_A},h_{n_{A}+1},\ldots,h_{n})$ and 
$(\Mpbe)_{1j}=(-g_1,\ldots,-g_{n_A},g_{n_{A}+1},\ldots,g_{n})$, respectively,
the coefficient of the  operators $\hat \mu_{\bfbe}$ and $\hat \nu_{\bfbe}$ to test entanglement in the seed
bipartition $\bfbe|\vec \beta$ are 
$(\Mxbe)_{1j}=(h_1,\ldots,h_{n_A},h_{n_{A}+1},\ldots,h_{n})$
and $(\Mpbe \bmath \Lambda_{\bfbe})_{1j}=(g_1,\ldots,g_{n_A},g_{n_{A}+1},\ldots,g_{n})$, respectively. Of course the condition that $\hat \mu_{\bfbe}$ and $\hat \nu_{\bfbe}$
conmute impose some restrictions on the possible coefficients $h_j$ and $g_j$. 
\par
In order to test entanglement in the rest of the bipartitions of the same class $(n_A,n_B)$, we can use the matrices 
\beq
\label{N-matrices}
\Mpal=\bmath{P}_{\bfbe\vec \alpha }\Mpbe \bmath{P}^T_{\bfbe \vec \alpha}\;\;,\;\;
\Mxal=\gamma_{\bfbe}\bmath{P}_{\bfbe\vec \alpha }(\Mpbe^{-1})^T \bmath{P}^T_{\bfbe \vec \alpha}.
\eeq
One can check that these also verify Eq. \eqref{matrixMbfal}.
This means that the first rows of the seed matrices $\Mxbe$ and $\Mpbe$ were scrambled 
by the matrix $\bmath{P}^T_{\bfbe \vec \alpha}$ and then mapped to the $\alpha_1$ row by the matrix $\bmath{P}_{\bfbe \vec \alpha}$.
Thus, from \eqref{new-seed-operators} the non-local operators to test bipartition $\vec \alpha|\vec \beta$ are:
\bse
\label{new-operators-l1}
\bea
\hat {{\mu}}_{\alpha_1,\bfal}&=&\sum_{j=1}^n(\Mxbe \bmath{P}^T_{\bfbe \vec \alpha})_{\alpha_1,j}\hat x_j\\
\hat {{\nu}}_{\alpha_1,\bfal}&=&\pm \sum_{j=1}^n(\Mpbe \bmath \Lambda_{\bfbe}
 \bmath{P}^T_{\bfbe \vec \alpha})_{\alpha_1,j}\hat x_j.
\eea
\ese
We emphasize that although the mirrored operators 
\beq
\label{old-operators-seed-induced}
\hat u_{\alpha_1,\bfal}=\hat {{\mu}}_{\alpha_1,\bfal}\;\;\mbox{and}\;\;
\hat { v}_{\alpha_1,\bfal}=\sum_{l=1}(\Mpbe \bmath {P}^T_{\bfbe \vec \alpha})_{\alpha_1,l}\hat x_l,
\eeq 
are  different from the mirrored operators in Eq.(\ref{old-operators-seed}) for the seed bipartition, 
their commutator $[\hat {u}_{\alpha_1,\bfal},\hat { v}_{\alpha_1,\bfal}]=
[\hat {u}_{1,\bfbe},\hat { v}_{1,\bfbe}]=i\gamma_{\bfbe}$ is the same.
So, using the matrices in Eqs.(\ref{N-matrices}), the lower bound in Eq.(\ref{genuine-same}) is equal for all bipartitions 
of the same class $(n_A,n_B)$. 
Thus, we can use this result to obtain a single 
pair of non-local operators to be used in the genuine multipartite  entanglement criterion in Eq.(\ref{genuine-criteria}). 
\par 
For an arbitrary  bipartition class $(n_A,n-n_A)$ ($1\leq n_A\leq n_A^{\mbox{\scriptsize max}}$) we set two type of seed matrices: $\tilde {\bmath M}_{p,\bfbe}$ and $\tilde {\bmath M}_{p,\bfbe}^{\prime}$ respectively.
For bipartitions $\vec \alpha|\vec \beta$ such that {$n\notin \vec \alpha$}, we choose the seed matrix $\tilde {\bmath M}_{p,\bfbe}$ with the structure given in Eq.\eqref{seed-matrix-form} of
Appendix \ref{Appendix2}. 
{It's first row} is: 
\beq
\label{first-seed-matrix}
(\tilde {\bmath M}_{p,\bfbe})_{1j}=(- g,\ldots,-g,g\ldots,g,g^\prime),
\eeq
where the minus sign appears in the first $n_A$ positions, 
$g=-\gamma/2h$ and $g^\prime=\gamma(n-1)/2h^\prime$
($h,h^{\prime},\gamma$ arbitrary real numbers). 
For bipartitions such that {$n\in \vec \alpha$}, we choose the seed matrix 
$\tilde {\bmath M}_{p,\bfbe}^{\prime}$ whose first row is:
\beq
\label{second-seed-matrix}
(\tilde {\bmath M}_{p,\bfbe}^\prime)_{1j}=(-g,\ldots,-g^\prime,\ldots,- g,\ldots,-g,g,\ldots, g),
\eeq
where  again the minus sign stands in the first $n_A$ positions and $g^\prime$ is located in the position
$i$ ($1\leq i\leq n_A$) such that $\alpha_i=n$ with 
$\alpha_i$ a component of $\vec \alpha$. 
Therefore,  because it is always true that $\Mxbe=\gamma_{\bfbe} (\Mpbe^{-1})^T$,
for the x-matrix associated with the p-matrix in Eq.(\ref{first-seed-matrix}) 
({\it i.e.} when $n\notin \vec \alpha$) we 
have,
\beq
\label{first-seed-matrix-X}
(\tilde{\bmath M}_{x,\bfbe})_{1j}=(h,h,\ldots,h,h^\prime).
\eeq 
And for the x-matrix associated with the p-matrix in Eq.(\ref{second-seed-matrix}),
({\it i.e.} when $n \in \vec \alpha$) we have,
\beq
\label{second-seed-matrix-X}
(\tilde{\bmath M}_{x,\bfbe}^\prime)_{1j}=(h,\ldots,h^\prime,h,\ldots,h),
\eeq
where the location of $h^\prime$ is in the position
$i$ ($1\leq i\leq n_A$) such that $\alpha_i=n$. 
These seed matrices set the {mirrored} operators in Eqs.(\ref{old-operators-seed}) and (\ref{old-operators-seed-induced})  with
the commutator,
\beq
\label{commutator-old-ope-todas-bipart}
[\hat u_{l_1,\bfal},\hat {v}_{l_1,\bfal}]=[\hat u_{1,\bfbe},\hat {v}_{1,\bfbe}]=i 
n_A\gamma,
\eeq 
if the mode $n\notin \vec \alpha$ 
and 
\beq
\label{commutator-old-ope-todas-bipart-2}
[\hat u_{l_1,\bfbe},\hat {v}_{l_1,\bfbe}]=[\hat u_{1,\bfbe},\hat {v}_{1,\bfbe}]=
-i(n-n_A)\gamma,
\eeq
if the mode $n\in \vec \alpha$. 
Then, because $(\tilde{\bmath M}_{p,\bfbe}^\prime\bmath\bmath \Lambda_{\bfbe} 
\bmath{P}^T_{\bfbe \vec \alpha})_{\alpha_1j}=
(\tilde{\bmath M}_{p,\bfbe}\bmath\bmath \Lambda_{\bfbe} \bmath{P}^T_{\bfbe \vec \alpha})_{\alpha_1j}=
(g,\ldots,g,g^\prime)$ and 
$(\tilde{\bmath M}_{x,\bfbe}^\prime \bmath{P}^T_{\bfbe \vec \alpha})_{\alpha_1j}=
(\tilde{\bmath M}_{x,\bfbe} \bmath{P}^T_{\bfbe \vec \alpha})_{\alpha_1j}=
(h,\ldots,h,h^\prime)$,
the commuting operators in Eqs.(\ref{new-seed-operators}) and (\ref{new-operators-l1})
to test entanglement in all the bipartition of the 
class $(n_A,n-n_A)$ are equal to:
\bse
\label{single-operators}
\bea
\hat {{\mu}}&=&h \hat x_1+h \hat x_2+\ldots+h \hat x_{n-1}+h^\prime\hat x_n\\ 
\hat {{\nu}}&=&\frac{\gamma}{2}\left(-\frac{\hat p_1}{h}-\ldots-
\frac{\hat p_{n-1}}{h}+\frac{(n-1)\hat p_n}{h^\prime}\right),
\eea
\ese
with  $[\hat \mu,\hat \nu]=0$.
This family of single pairs of operators are the ones that must be used in our genuine entanglement criterion in  Eq.(\ref{genuine-criteria}) with $\gamma_{\mbox{\scriptsize min}}=\mbox{ min}_{\{\vec \alpha\}} \left\{|\gamma_{\vec \alpha}|\right\}= \mbox{min}_{n_A}\left\{ n_A|\gamma|,(n-n_A)|\gamma|\right\}=|\gamma|$,
where the minimization is over the values 
$1\leq n_A\leq n_A^{\mbox{\scriptsize max}}$.   The three free parameters of this family ($h, h^\prime, \gamma$) 
increases the chances to detect genuine entanglement in a $n$-mode systems using the single inequality in Eq.(\ref{genuine-criteria}).

\par
We recover the 
commuting non-local operators that appear in Eq.(30) of \cite{vanloock03}, if we relabel mode $1$ as $n$ and set
$h^\prime=1$ and  $h=-1/\sqrt{n-1}$, then $\gamma=2/(n-1)$. Here  
the lower bound must be given by $\gamma=2/(n-1)$ according to 
our Eq.(\ref{genuine-criteria}) with $F=F_{Lin}$ and $f_{Lin}(|\gamma|) = |\gamma|=2/(n-1)$.
{The factor of $1/2$ of the lower bound  in  Ref. \cite{vanloock03} is due to the fact that there the commutation relation is $[\hat x_j,\hat p_k]=i\delta_{jk}/2$ instead of $[\hat x_j,\hat p_k]=i\delta_{jk}$, as defined here.}
This proves that the popular entanglement criteria in  \cite{vanloock03} is indeed a PPT criteria. Although, in  \cite{vanloock03}
it was not proven that the single inequality in our Eq.(\ref{genuine-criteria}) with $F=F_{Lin}$ and $f_{Lin}(|\gamma|) = |\gamma|=2/(n-1)$ is never violated by biseparable states $\hat \rho_{bs}$,
this was pointed out in \cite{reid13}. 
However, {here we have gone a step further by proving that not only the sum of variance UR, but in fact 
any UR of the form Eq.(\ref{UR}) can be used to set a genuine entanglement criteria 
with \eqref{genuine-criteria} for any pair of linear non-local operators of the family \eqref{single-operators}.} 
In this regard, we stress that with similar experimental effort one can test entropic 
entanglement criteria, which outperform variance criteria in general. 
{It is worth noting that the genuine tripartite entanglement criteria in Eq.(1) of \cite{Armstrong2015}
is a special case  of our genuine criterion given in \eqref{genuine-criteria}
with $F=F_{H}$ and $f_{H}(|\gamma|) = |\gamma|^2/4$ where the single 
pair of operators used $\mu=\hat x_1-(x_2+x_3)/\sqrt{2}$ and $\nu=p_1+(p_2+p_3)/\sqrt{2}$ 
can be mapped in our family of commuting operators in \eqref{single-operators} 
if we rename mode $1$ as mode $3$ and set $h=-1/\sqrt{2}$ and $h^{\prime}=\gamma=1$.
 The lower bound  in  Ref. \cite{Armstrong2015} is 1 instead of 
 $f_{H}(|\gamma|=1) = 1/4$ of our case 
 due to the fact that there the commutation relation is $[\hat x_j,\hat p_k]=2i\delta_{jk}$ instead of $[\hat x_j,\hat p_k]=i\delta_{jk}$, as defined here. }
\par 
A limitation of using the single inequality in Eq.(\ref{genuine-criteria}) as 
a genuine entanglement criterion is that the range of URs that can be used to set 
the functional $F$ (and the function $f$) is restricted
to those URs  of the form in Eq.(\ref{UR}) valid for arbitrary linear non-local observables, excluding the possibility to use 
URs valid for conjugate pairs, {\it i.e.}
for those pairs of operators related by a $\pi/2$ rotation. 
For example, we can not use the UR in Eq.(\ref{eqBirulapm}) to set the functional $F$
and the function $f$ in our PPT-URNLO genuine entanglement criterion in Eq.(\ref{genuine-criteria}).
This is because the pairs of
mirrored operators $(\hat u_{l_1,\bfal},\hat {v}_{l_1,\bfal})$ and $(\hat u_{1,\bfbe},\hat {v}_{1,\bfbe})$ 
associated with the  pairs  of operators $\hat \mu$ and $\hat \nu$ in Eq.(\ref{single-operators}) cannot be canonical conjugated. 
In order to see this, notice that the seed matrices $\tilde {\bmath M}_{p,\bfbe}$ and $\tilde {\bmath M}_{p,\bfbe}^\prime$, whose first rows are in Eqs.(\ref{first-seed-matrix}) and (\ref{second-seed-matrix}) respectively, and the matrizes $\tilde {\bmath M}_{x,\bfbe}$ and $\tilde {\bmath M}_{x,\bfbe}^\prime$, whose first rows are in and Eq.(\ref{first-seed-matrix-X}) and (\ref{second-seed-matrix-X}) respectively,
do not satisfied the condition in Eq.(\ref{matrixMbfal-orthogonal}), that guarantee the conjugation between the mirrored  operators, for any value of the parameters $h,h^{\prime},\gamma$.
We can fix this limitation if we consider PPT-URNLO entanglement criteria based on a set of pairs of commuting operators.
\subsection{Criterion with several pairs  of commuting non-local operators}
\label{sec:severalpairs}

It is also possible to have a genuine PPT-UNRLO entanglement criterion that consists also in a single inequality but, contrary to the one
in Eq.(\ref{genuine-criteria}), involves several pairs of commuting operators.  We start defining a set of pairs $\{(\hat{{\mu}}_m,\hat {{\nu}}_m)\}$ of commuting operators with $m=1,\ldots,M$ ($M\leq L$) of the form $\hat {{\mu}}_m=\sum_{j=1}^{n}h_{mj}\hat x_j$ and $\hat { {{\nu}}}_m=\sum_{j=1}^{n}g_{mj}\hat p_j$, where $h_{mj},g_{mj}$ are real numbers not all equal to zero.  We also define
$\gamma_{m,\vec \alpha}\equiv\sum_{j=1}^{n} \bar g_{mj} h_{mj}$ where  for all values of $m$, $\bar g_{mj}=-g_{mj}$ if  $j$ is one component of the vector $\vec \alpha$ or $\bar g_{mj}=g_{mj}$ otherwise. Now, for a given set of coefficients $h_{mj}$ and $g_{mj}$ ($m$ fixed and $j=1,\ldots,n$), we must check all the possible locations of the minus sign of $\bar g_{mj}$ within
$\gamma_{m,\vec \alpha}\equiv\sum_{j=1}^{n} \bar g_{mj} h_{mj}$ and see for which combinations of 
locations $\gamma_{m,\vec \alpha}$ is different from zero.  
According to  
Eq.(\ref{gamma-bfal}), for $m$ fixed, the number $\gamma_{m,\vec \alpha}$ is the commutator  of the operator pair $(\hat u_m,\hat v_m)$ where the location of the minus sign in $\gamma_{m,\vec \alpha}$ indicates which modes were partial transposed.  
\par
We denote  the  subset of all bipartitions $\{\vec \alpha|\vec \beta\}$ of the system where the values of $\gamma_{m,\vec \alpha}$ are not zero as $\{\vec \alpha\}_m$ (for every fixed value of
$m$). This means, that we can test entanglement with the pair $(\hat \mu_m,\hat \nu_m)$ in all the bipartitions $\{\vec \alpha\}_m$ 
with the single inequality (like the one in Eq.(\ref{genuine-criteria})), 
\beq
\label{bipartite-entag-test-several-operators}
F[\hat \rho,P_{\hat \mu_m}, P_{\hat \nu_m}]\geq
f(\gamma_{m,\mbox{\scriptsize min}})\geq 0,
\eeq
where we call $\gamma_{m,\mbox{\scriptsize min}}=\stackrel[\{\vec \alpha\}_m]{\mbox{ min}}{}\left\{|\gamma_{m,\vec \alpha}|\right\}>0$.
Notice, that for every pair of observables $(\hat \mu_m,\hat \nu_m)$ correspond a set $\{(\hat u_{m,l},\hat v_{m,l})\}$ of mirrored observables, with $l=1,\ldots,L_m$ and $L_m$ the total number of bipartitions in the set $\{\vec \alpha\}_m$. 
This means that for every bipartition in the set $\{\vec \alpha\}_m$, we have 
a different pair of mirrored observables $\hat u_{m,l}=\sum_{j=1}^n h_{m,j}\hat x_j$
and $\hat u_{m,l}=\sum_{j=1}^n \bar g_{m,j}\hat p_j$ (the index $l$ is counting the different 
location of minus sign such $\gamma_{m,\vec \alpha}\equiv\sum_{j=1}^{n} \bar g_{mj} h_{mj}\neq 0$).
\par
Now,  consider functionals $F$ such either i) are concave with respect to 
a convex sum of density operators or ii) verify $F>F^{\prime}$ with $F^{\prime}$ concave with respect to a convex sum of density operators.
Then, we can proceed like in Eq.(\ref{demostration-no-violation}) and 
recognise that inequality (\ref{bipartite-entag-test-several-operators}) is never violated by 
biseparable states $\hat \rho_{bs}$  with 
$p_{\{\vec \alpha|\vec \beta \}}=0$ corresponding to the bipartitions such that $\{\vec \alpha|\vec \beta \}\neq\{\vec \alpha\}_m$ (see Eq.(\ref{biseparable-states})).
However, states $\hat \rho_{bs}$ with $p_{\{\vec \alpha|\vec \beta \}}=0$ for $\{\vec \alpha|\vec \beta \}=\{\vec \alpha\}_m$
do not necessary verify the inequality (\ref{bipartite-entag-test-several-operators}).
Furthermore, inequality (\ref{bipartite-entag-test-several-operators}) only tests bipartite entanglement 
in the bipartitions of the set $\{\vec \alpha\}_m$ ($m$ fixed) that does not necessarily correspond to all the bipartitions of the system.
But, if  all of the sets $\{\{\vec \alpha\}_m, m=1,\ldots,M\}$ cover all possible bipartitions of the system (with possible repetitions), then all biseparable states
$\hat \rho=\hat \rho_{bs}$ satisfy:
\beq
\label{last-criterium}
\sum_{m=1}^M F[\hat \rho,P_{\hat {{\mu}}_m}(\xi),P_{\hat{\nu}_m}(\xi)]\geq
\lfloor\Theta\rfloor\;f\left(
\tilde \gamma_{\mbox{\scriptsize min}}
\right)\geq f\left(
\tilde \gamma_{\mbox{\scriptsize min}}
\right),
\eeq
with
$\tilde\gamma_{\mbox{\scriptsize min}}=\stackrel[m]{\mbox{ min}}{}\left\{\gamma_{m,\mbox{\scriptsize min}}\right\}$
and $\lfloor\Theta\rfloor= \lfloor \sum_{m=1}^M
\sum_{\{\vec \alpha\}_m}p_{\{\vec \alpha\}_m}\rfloor \geq 1$, where $\lfloor x \rfloor$ is the largest integer not greater than $x$. 
Therefore, Eq.(\ref{last-criterium}) constitutes a PPT-UNRLO entanglement criterion for genuine $n$-partite entanglement.
  
\par
Consider for example, in the case of $4$-mode states, 
the set of  commuting operators considered in Eq.(39) of Ref.  
\cite{vanloock03} (see also \cite{reid13}):
\bea
\hat \mu_1&=&\hat x_1-\hat x_2\;\;,\;\;
\hat \nu_1=\hat p_1+\hat  p_2+g_{13} \hat p_3+ g_{14} \hat p_4,\nonumber \\
\hat \mu_2&=&\hat x_2-\hat x_3\;\;,\;\;
\hat \nu_2=g_{21}\hat p_1 + \hat p_2+\hat p_3 +g_{24} \hat p_4,
\label{six-operators}\\
\hat \mu_3&=&\hat x_1-\hat x_3\;\;,\;\;
\hat \nu_3=\hat p_1 +g_{32} \hat p_2+\hat p_3 +g_{34} \hat p_4,\nonumber
\\
\hat \mu_4&=&\hat x_3-\hat x_4\;\;,\;\;
\hat \nu_4=g_{41}\hat p_1 +g_{42} \hat p_2+\hat p_3 + \hat p_4,\nonumber\\
\hat \mu_5&=&\hat x_2-\hat x_4\;\;,\;\;
\hat \nu_5=g_{51}\hat p_1 + \hat p_2+g_{53}\hat p_3 +\hat p_4,\nonumber\\
\hat \mu_6&=&\hat x_1-\hat x_4\;\;,\;\;
\hat \nu_6=\hat p_1 +g_{62} \hat p_2+g_{63}\hat p_3 + \hat p_4. \nonumber
\eea
For the first pair we have that $\gamma_{1,\vec \alpha}\equiv
\sum_{j=1}^{4} \bar g_{1j} h_{1j}\neq0$ (where $g_{11}=g_{22}=h_{11}=1$, $h_{12}=-1$ and $h_{13}=h_{14}=0$) only when we consider partial transposition in the bipartitions:
$\{\vec \alpha\}_1=\{1|234,2|134,13|24,14|23\}$. Therefore, with the pair $(\hat \mu_1,\hat \nu_1)$ we can test bipartite entanglement in the bipartitions in the set $\{\vec \alpha\}_1$ using the inequality 
in Eq.(\ref{bipartite-entag-test-several-operators}) with  $\gamma_{1,\mbox{\scriptsize min}}=2$, 
since $\gamma_{1,\vec \alpha}=2$ for all the bipartitions in the set $\{\vec \alpha\}_1$. 
Equivalently we can test bipartite entanglement with the rest of the pairs of operators in the following bipartitions: $\{\vec \alpha\}_2=\{2|134, 3|124, 12|34,13|24\}$, $\{\vec \alpha\}_3=\{1|234,3|124,12|34,14|23\}$, $\{\vec \alpha\}_4=\{3|124,4|123,13|24,\\14|23\}$, $\{\vec \alpha\}_5=\{2|134,4|123,12|34,  14|23\}$,
$\{\vec \alpha\}_6=\{1|234,4|123,12|34,13|24\}$ (
where $\gamma_{m,\vec \alpha}=2$ 
so $\tilde\gamma_{\mbox{\scriptsize min}}=\gamma_{m,\mbox{\scriptsize min}}=2$ with $m=1,\ldots,6$). Also,
$\lfloor\Theta\rfloor=\lfloor3 (p_{1|234}+p_{2|134}+p_{3|124}+p_{4|123})+4(
p_{12|34}+p_{13|24}+p_{14|23})\rfloor=3$ so we can use the lower bound $3f\left(
\tilde \gamma_{\mbox{\scriptsize min}}
\right)$
in our genuine entanglement criterion in Eq.(\ref{last-criterium}).
In the case when the functional is $F=F_{Lin}$ (and therefore $f(\tilde\gamma_{\mbox{\scriptsize min}}) =f_{Lin}(\tilde\gamma_{\mbox{\scriptsize min}}) =|\tilde\gamma_{\mbox{\scriptsize min}}|=2$) we recover the genuine entanglement criterion given in 
Eq.(43) of \cite{reid13}.  {Again, the difference in the lower bound comes from  the difference in the canonical commutation relation that they used.} We stress that the set of inequalities in Eq.(39) of Ref. \cite{vanloock03},
that correspond to our inequalities in Eq.(\ref{bipartite-entag-test-several-operators}) with the operators 
in Eq.(\ref{six-operators}) and $F=F_{Lin}$ ($f_{Lin}(\gamma_{m,\mbox{\scriptsize min}}) =|\gamma_{m,\mbox{\scriptsize min}}|=2$), do not constitute a genuine multipartite entanglement criterion, as was pointed out in \cite{reid13}. Only when the single inequality in Eq.(\ref{last-criterium}) is used, a genuine multipartite entanglement criterion is achieved.
\par 
Also, we can easily recover the four-partite genuine entanglement criterion given in Eq.(44) of \cite{reid13}
if we use the set of pairs of non-local operators $\{(\hat \mu_m,\hat \nu_m)\}$ ($m=1,2$)
where the pair $(\hat \mu_2,\hat \nu_2)$ is the one already given in our Eq.(\ref{six-operators}), and the new commuting pair is
$(\hat \mu_1=\hat x_1-\hat x_4-(\hat x_2+\hat x_3),\hat \nu_1=\hat p_1-\hat p_4+\hat p_2+\hat p_3)$.  With this new pair we can test bipartite entanglement in the bipartitions 
$\{\vec \alpha\}_1=\{1|234,2|134,3|124,4|123,14|23\}$, with $|\gamma_{1,1|234}|=|\gamma_{1,2|134}|=|\gamma_{1,3|124}|=|\gamma_{1,4|123}|=2$ and $|\gamma_{1,14|23}|=4$ respectively, so 
$\gamma_{1,\mbox{\scriptsize min}}=2$ and
$\tilde\gamma_{\mbox{\scriptsize min}}=\mbox{ min}\left\{\gamma_{1,\mbox{\scriptsize min}},\gamma_{2,\mbox{\scriptsize min}}\right\}=2$.
 In this case we have 
 $\lfloor\Theta\rfloor=\lfloor p_{1|234}+2 p_{2|134}+2 p_{3|124}+p_{4|123}+
p_{12|34}+p_{13|24}+p_{14|23}\rfloor=2$.
We do not develop explicitly all cases but 
it is straightforward to verify that all genuine multipartite entanglement criteria presented in \cite{reid13} can be recovered with the systematic presented in this work.
The same is true for all the genuine $3$-mode genuine entanglement criteria presented in 
\cite{Shalm2012}. For example, the genuine entanglement criteria in Eq.(5) of \cite{Shalm2012} 
is a special case of our genuine PPT-UNRLO entanglement criterion
in Eq.(\ref{last-criterium}) with $F=F_H$ (and $f(\tilde\gamma_{\mbox{\scriptsize min}})=f_H(\tilde\gamma_{\mbox{\scriptsize min}})= 
1/4|\tilde\gamma_{\mbox{\scriptsize min}}|^2$). 
In this case we only need two pairs of non-local observables:
\bea
\hat \mu_1&=&\hat x_1-\hat x_2\;\;,\;\;
\hat \nu_1=\hat p_1+\hat  p_2 +\hat p_3,\nonumber \\
\hat \mu_2&=&\hat x_1-\hat x_3\;\;,\;\;
\hat \nu_2=\hat p_1+\hat  p_2 +\hat p_3.
\eea
With the first pair $(\hat \mu_1,\hat \nu_1)$ we can test bipartite entanglement in the bipartitions
$\{\vec \alpha\}_1=\{1/23,2/13\}$, and with the second pair  in the bipartitions $\{\vec \alpha\}_2=\{1/23,3/12\}$,
where $|\gamma_{m,\vec \alpha}|=2=\gamma_{m,\mbox{\scriptsize min}}$ ($m=1,2$) for all the bipartitions,
and therefore $\tilde\gamma_{\mbox{\scriptsize min}}=2$. Also,
 $\lfloor\Theta\rfloor=\lfloor 2p_{1|23}+ p_{2|13}+p_{3|12}\rfloor=1$. 

\par
Our systematic {approach} also unveils many alternatives to construct genuine entanglement criteria 
for the same set of  commuting non-local observables. This come through the possibility to use any UR relation of the form in Eq.(\ref{UR}) valid for the associated set of mirrored non-local observables $\{(\hat{{u}}_{m,l},\hat {{v}}_{m,l})\}$.  
In this regards it is worth noting that it is always possible to choose 
the set  $\{(\hat{{\mu}}_m,\hat {{\nu}}_m)\}$ ($m=1,\ldots,M\leq L$) in such a way that, for every  pair $(\hat{{\mu}}_m,\hat {{\nu}}_m)$, the associated set $\{(\hat{{u}}_{m,l},\hat {{v}}_{m,l})\}$ consists of conjugate pairs.  So, we can also use in our entanglement criterion in 
Eq.(\ref{last-criterium}), any UR of the form in Eq.(\ref{UR}) that are valid only for conjugate pairs of observables, such as those involving the R\'enyi entropy 
Eq.(\ref{eqBirulapm}). It is instructive to develop an example.
In a $4$-mode system, we can use the set of  observables:
\bea
\nonumber \\
\hat \mu_1&=&\frac{\hat x_1-\hat x_2}{\sqrt{2}}\;\;,\;\;
\hat \nu_1=\frac{\hat p_1+\hat  p_2}{\sqrt{2}},\nonumber \\
\hat \mu_2&=&\frac{\hat x_2-\hat x_3}{\sqrt{2}}\;\;,\;\;
\hat \nu_2= \frac{\hat p_2+\hat p_3}{\sqrt{2}},
\label{six-operators-2}\\
\hat \mu_3&=&\frac{\hat x_1-\hat x_3}{\sqrt{2}}\;\;,\;\;
\hat \nu_3=\frac{\hat p_1 +\hat p_3}{\sqrt{2}},\nonumber
\\
\hat \mu_4&=&\frac{\hat x_3-\hat x_4}{\sqrt{2}}\;\;,\;\;
\hat \nu_4=\frac{\hat p_3 + \hat p_4}{\sqrt{2}},\nonumber\\
\hat \mu_5&=&\frac{\hat x_2-\hat x_4}{\sqrt{2}}\;\;,\;\;
\hat \nu_5= \frac{\hat p_2+\hat p_4}{\sqrt{2}},\nonumber\\
\hat \mu_6&=&\frac{\hat x_1-\hat x_4}{\sqrt{2}}\;\;,\;\;
\hat \nu_6=\frac{\hat p_1 + \hat p_4}{\sqrt{2}}. \nonumber
\eea
The sets of bipartitions $\{\vec \alpha\}_m$  that these operators test for bipartite entanglement coincide with the sets of bipartitions that the operators in Eq.(\ref{six-operators})  test for bipartite entanglement. In this case, $|\gamma_{m,\vec \alpha}|=\gamma_{m,\mbox{\scriptsize min}}=\tilde\gamma_{\mbox{\scriptsize min}}=1$ with $m=1,\ldots,6$.
The associated set of  mirrored non-local observables $\{(\hat{{u}}_{m,l},\hat {{v}}_{m,l})\}$,
with $m=1,\ldots,6$ and $l=1,\ldots,4$, are all conjugate pairs.
Indeed, according to Eq.(\ref{old-operators}), for each conjugate pair $(\hat{{u}}_{m,l},\hat {{v}}_{m,l})$ we have associated  matrices $\bmath M_{x,\bfal=\vec \alpha}$ and $\bmath M_{p,\bfal=\vec \alpha}$
whose first rows, for example, 
correspond to the coefficients of the operators $\hat{{u}}_{m,l}$ and $\hat{{v}}_{m,l}$
respectively.
In this case,  because $\hat{{u}}_{m,l}$ and $\hat{{v}}_{m,l}$ are conjugate pairs, 
the different matrices satisfy  $\Mxal=\gamma_{m,\vec \alpha} (\Mpal)$, as we can readily check.
\par
\section{Example}
\begin{figure}
\includegraphics[width=8cm]{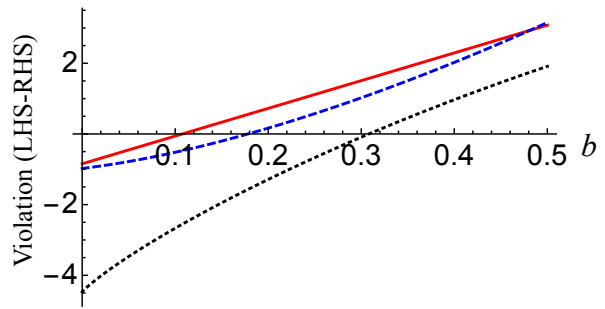}
\caption{(Color online.) Violation of three entanglement criteria for state \eqref{eq:examplestate} as a function of the mixing parameter $b$ (both are dimensionless quantities).  Genuine quadripartite entanglement is identified for negative values.  The red solid curve corresponds to the linear criteria \eqref{eq:lin}, the blue dashed line to the product criteria \eqref{eq:prod}, and the black dotted line to the entropic criteria \eqref{eq:ent}.  For this non-gaussian state, the entropic criteria detects entanglement in regions where the variance criteria fail.}
\label{fig:example}
\end{figure}

The utility of our technique can be better appreciated by an example.  Let us consider the quadripartite state
\begin{equation}
\hat \rho = (1-b) \ket{\psi}\bra{\psi} + b \ket{\mathrm{vac}}\bra{\mathrm{vac}},
\label{eq:examplestate}
\end{equation}
where $\ket{\mathrm{vac}}$ is the four-mode vacuum state and the state
\begin{align}
 \ket{\psi} = & \iiiint   \frac{1} {\pi  \sqrt{s^2 t^2}}e^{-\frac{\left({x_1}+\frac{{x_2}+{x_3}}{\sqrt{2}}\right)^2}{4 s^2}} e^{-\frac{\left(\frac{{x_3}-{x_2}}{\sqrt{2}}+{x_4}\right)^2}{4
   s^2}} \times \\ \nonumber 
 & e^{-\frac{\left({x_1}-\frac{{x_2}+{x_3}}{\sqrt{2}}\right)^2}{4 t^2}}e^{-\frac{\left(\frac{{x_3}-{x_2}}{\sqrt{2}}-{x_4}\right)^2}{4 t^2}} \ket{x_1}\ket{x_2}\ket{x_3}\ket{x_4}
   \end{align}
  can be {produced} by creating two-mode squeezed states (modes 1/2 and 3/4) and then combining modes 2 and 3 on a 50/50 beam splitter \cite{braunstein05}.  Here the variables $s$ and $t$ are related to the usual squeezing parameter $r$ by $s=(e^r)/2$ and $t=(e^{-r})/2$, where $r\longrightarrow \infty$ corresponds to infinite squeezing.  The state $\rho$ is an incoherent combination of the state $\ket{\psi}$ and the vacuum state.  We will probe the entanglement using the operators 
  \begin{subequations}
  \begin{align}
  \hat{\mu}_1 & = \hat{x}_1-\frac{{\hat{x}_2}+{\hat{x}_3}}{\sqrt{2}}, \\
  \hat{\nu}_1 & = {\hat{p}_1}+\frac{{\hat{p}_2}+{\hat{p}_3}}{\sqrt{2}}, \\ 
 \hat{\mu}_2 & = \frac{{\hat{x}_3}-{\hat{x}_2}}{\sqrt{2}}-{\hat{x}_4}, \\
  \hat{\nu}_2 & = \frac{{\hat{p}_3}-{\hat{p}_2}}{\sqrt{2}}+{\hat{p}_4}. 
 \end{align}
  \end{subequations}
The marginal probability distributions associated to these operators can either be measured directly or obtained as marginals of the  probability distributions $P(x_1,x_2,x_3,x_4)$ and $P(p_1,p_2,p_3,p_4)$.  Following Table \ref{tab:2}, there are seven possible bipartitions. One can see immediately that the operator pair $\hat{\mu}_1$ and $\hat{\nu}_1$ could detect entanglement in every possible bipartition,
{except  in the bipartition $4|123$, using the type of criterion in Eq.(\ref{bipartite-entag-test-several-operators}). Analogously, the pair $\hat{\mu}_2$ and $\hat{\nu}_2$ could detect entanglement in all the bipartitions except in the bipartition $1|234$}.  Thus, following the procedure outlined in section \ref{sec:severalpairs}, we can use criteria \eqref{last-criterium} with $M=2$, and also with the knowledge that each possible bipartion appears at least once in the sum over all tested bipartitions in the argument of the $\Theta$ function.  Thus, we have $\lfloor \Theta\rfloor =1$.  Furthermore, direct calculation of the commutators $\gamma_{m,\vec{\alpha}}$ for all tested bipartitions (parameterized by $\vec{\alpha}$) shows that $\gamma_{m,\vec{\alpha}} \geq 1$, so that $\tilde{\gamma}_\mathrm{min}=1$.  We can then test the three entanglement criteria provided by the set of inequalities \eqref{all-inequalities} in the multipartite form given by Eq. \eqref{last-criterium}. Specifically, we test the linear inequality:
\begin{equation}
\Delta\hat{\mu}_1^2 + \Delta\hat{\nu}_1^2+\Delta\hat{\mu}_2^2+\Delta\hat{\nu}_2^2 \geq 1,
\label{eq:lin}
 \end{equation}
 the product inequality:
 \begin{equation}
2 \Delta\hat{\mu}_1\Delta\hat{\nu}_1+2\Delta\hat{\mu}_2\Delta\hat{\nu}_2 \geq 1,
\label{eq:prod}
 \end{equation}
and the entropic inequality:
\begin{equation}
h[P_{\hat \mu_1}]+h[P_{\hat \nu_1}]+h[P_{\hat \mu_2}]+h[P_{\hat \nu_2}] \geq \ln(\pi e). 
\label{eq:ent}
 \end{equation}
Violation of any of these inequalities guarantees genuine quadripartite entanglement.  Figure \ref{fig:example} shows the violation of these inequalities for the state \eqref{eq:examplestate}  as a function of the mixing parameter $b$.  Here we chose squeezing parameter $r=2$.  One can see that the entropic criteria detects entanglement in regions where both the linear variance and variance product criteria fail.   {We emphasise that if the joint distribution probabilities, $P(x_1,x_2,x_3,x_4)$ and $P(p_1,p_2,p_3,p_4)$, were experimentally sampled, we have the freedom to choose any set $\{(\hat \mu_m(x_1,x_2,x_3,x_4),\hat \nu_m(p_1,p_2,p_3,p_4))\}$ to test genuine multipartite entanglement,
using any uncertainty relation in criterion \eqref{last-criterium} that depends only on the 
marginal distributions $P_{\hat \mu_m}$ and $P_{\hat \nu_m}$.
 }
 \section{Final Remarks}
\label{SectionVI}
The detection of genuine multipartite entanglement is a necessary and important step in the realization of quantum information tasks that exploit correlations between many parties. In the case of bipartite entanglement of continuous variable systems, the most widely adopted entanglement criteria are those based on constraints provided by uncertainty relations on non-local operators. Here we have provided a general framework to construct these types of criteria, based on the positive partial transpose criteria and uncertainty relations for non-local operators.  Our criteria employ arbitrary uncertainty relations and consider bipartions of generic size.  We then use these results to build genuine multipartite entanglement criteria, and explicitly provide two categories.  The first allows one to identify genuine multipartite entanglment with the measurement of a single pair of operators, and the second allows one to perform measurements on subsets of the constituent systems.  These criteria are easily computable and experimentally friendly, in the sense that they require the reconstruction of a limited number of joint probability distributions.  We expect our results to be useful in identifying genuine entanglement in a number of experimental systems.


\begin{acknowledgements}
We acknowledge financial support from the Brazilian agencies FAPERJ, CNPq,  CAPES and the INCT-Informa\c{c}\~ao Qu\^antica. 
\end{acknowledgements}

\appendix 

\section{}
\label{Appendix1}

Here we are going to find the coefficients $(\Mxal)_{k,j}$ 
 and $(\Mpal)_{k,j}$ that define the mirrored 
 non-local operators in (\ref{old-operators}) such that the equality 
 $F[\hat \rho^{T_{\bfal}},P_{\hat u_{\bfal}},P_{\hat v_{\bfal}}]=F[\hat \rho,P_{\hat \mu_{\bfal}},P_{\hat \nu_{\bfal}}]$ in Eq.(\ref{criterion-general3-rewrite}) holds for any functional.
In order to do this, first note that the probability distributions $P_{\hat {\xi}}$ ($\xi =\mu,\nu$) are marginals:
\bea
P_{\hat {\xi}}=\int d\xi^\prime \; \tilde W_{T_\bfal}(\mu_{\bfal},\nu_{\bfal}), 
\eea
(where $\hat \xi=\hat \nu$ if $\xi^\prime=\nu$ and $\hat \xi=\hat \mu$ if $\xi^\prime=\mu$)
 of the marginal distribution
\bea
\tilde W_{T_\bfal}(u_{\bfal},v_{\bfal})&\equiv& \int 
\frac{d{\vectg{u}_{\bfal}} d{\vectg{v}}_{\bfal}}{|\gamma_{\bfal}|^{n}}
W_{T_{\bfal}}\left(\Mxal^{-1}\vectg{{u}}_{\bfal}, \Mpal^{-1}\vectg{{v}}_{\bfal}\right)\times\nonumber\\
&&\times\delta({ u}_{k,\bfal}-u_{\bfal})\delta({ v}_{k,\bfal}-v_{\bfal}) \nonumber 
\eea
of the Wigner  function $W_{T_{\bfal}}\left(\bsy {x},\bsy {p}\right)$ of the operator $\hat \rho^{T_{\bfal}}$.
Now, we remember that \cite{simon00}:
\beq
\label{wigner-transposition-condition}
W_{T_{\bfal}}(\vectg x,\vectg p)=W(\vectg x,\bmath \Lambda_{\bfal} \vectg p),\nonumber
\eeq
where $W$ is the Wigner function of the original state $\hat \rho$ and $\bmath \Lambda_{\bfal}$ 
is a diagonal matrix with ones in the location of modes that
are not transposed and negative ones in the location of modes that are transposed.
Making the change of variables 
$ \vectg{\mu}_{\bfal}=\Mmual \vectg{u}_{\bfal}$ and
 $\vectg{\nu}_{\bfal}=\Mnual \vectg{v}_{\bfal}$
 with the Jacobian equal to one we can write:
\bea
\lefteqn{\tilde W_{T_\bfal}(u_{\bfal},v_{\bfal})=}\nonumber\\
&&\int 
\frac{d{\vectg{{\mu}}_{\bfal}} d{\vectg{{\nu}}}_{\bfal}}{|\gamma_{\bfal}|^{n}}\, 
W\left(\Mxal^{-1}
\Mmual^{-1}
\vectg{\mu}_{\bfal} \;, \bmath \Lambda_{\bfal} \Mpal^{-1} \Mnual^{-1}\vectg{\nu}_{\bfal}\right)\times\nonumber\\
&&\times \delta\left(\sum_{i=1}^{n}{(\Mmual^{-1})_{ki} \mu}_{i,\bfal}-\mu_{\bfal} \right)\times\nonumber \\
&&\times \delta\left(\sum_{i=1}^{n}{(\Mnual^{-1})_{ki} \nu}_{i,\bfal}-\nu_{\bfal} \right)=
\tilde W(\mu_{\bfal},\nu_{\bfal}).
\label{demostration}
\eea
In order for $\tilde W(\mu_{\bfal},\nu_{\bfal})$ to be the marginal distribution of the 
Wigner function of the original state $\hat \rho$,
associated with operators $\hat \mu_{\bfal}$ and $\hat \nu_{\bfal}$, 
the following conditions must be fulfilled:
$\Mxal^{-1}\Mmual^{-1}=\Mxal^{-1}$   and
$\bmath \Lambda_{\bfal} \Mpal^{-1} \Mnual^{-1}=\Mpal^{-1}$.
This is equivalent to:
\bse
\label{ccond1}
\bea
\Mmual&=&\bmath 1\\
\Mnual &=&\Mpal \bmath \Lambda_{\bfal}\Mpal^{-1}=\Mpal \bmath \Lambda_{\bfal}\Mxal^{T}\gamma_{\bfal}^{-1},
\eea
\ese
with $\det(\Mnual)=\det\left(\bmath \Lambda_{\bfal}\right)=(-1)^{\tilde n}$ 
where $\tilde n=n_A$ for $\bfal=\vec \alpha$ and
$\tilde n=n_B=n-n_A$ when $\bfal=\vec \beta$ (thus the Jacobian in the change of variables in 
Eq.(\ref{demostration}) were indeed equal to one).
Note that 
$\Mnual$ is an involutory matrix, {\it i.e.},  $\Mnual^2 = \bmath 1 $ with signature $\tilde n$ (the signature is the number of elements equal to $-1$ in $\bmath \Lambda_{\bfal}$
\cite{levine1962}).
If conditions in Eqs.(\ref{ccond1}) are fulfilled then for the marginals $P_{\hat \mu_{\bfal}}$ and $P_{\hat \nu_{\bfal}}$ of $W(\mu_{\bfal},\nu_{\bfal})$ we have
\beq
\label{equality-P-marginals}
P_{\hat u_{\bfal}}=P_{\hat \mu_{\bfal}}\;\;\mbox{and}\;\;\;P_{\hat v_{\bfal}}=P_{\hat \nu_{\bfal}},
\eeq
where $P_{\hat u_{\bfal}}$ and $P_{\hat v_{\bfal}}$ are the marginals of $\tilde W_{T_\bfal}(u_{\bfal},v_{\bfal})$.
\par
The involutory property of $\Mnual$ ({\it i.e.} $\Mnual=\Mnual^{-1}$)
and the Eqs.(\ref{ccond1}) and (\ref{matrixMbfal}) allows us to recognise in (\ref{demostration}) 
the  operators $\hat \mu_{\bfal}$  and $\hat \nu_{\bfal}$ in Eq.(\ref{new-operators}) as:
\bse
\label{new-operators-bis-app}
\bea
\hat \mu_{\bfal} \equiv\hat {{\mu}}_{k,\bfal}&=&
\sum_{j=1}^n\gamma_{\bfal} ((\Mpal^{-1})^T)_{k,j}\hat x_j\\
\hat \nu_{\bfal}\equiv\hat\nu_{k,\bfal}&=&
\sum_{j=1}^{n}\;(\Mpal \bmath \Lambda_{\bfal})_{kj} \hat p_j,
\eea
\ese
where we use that $\Mxal=\gamma_{\bfal}(\Mpal^{-1})^T$. Therefore, comparing with Eqs.(\ref{new-operators}) we arrive at
\bse
\bea
\frac{ h_j}{\gamma_{\bfal} }&=&((\Mpal^{-1})^T)_{k,j}
  \label{hj}
   \label{hj-gj-a}\\
 g_j&=&(\Mpal \bmath \Lambda_{\bfal})_{kj},
 \label{hj-gj}
 \eea
 \ese 
 with $[\hat{\mu}_{\bfal},\hat\nu_{\bfal}]=i(\Mpal \bmath \Lambda_{\bfal}\Mxal^{T})_{kk}=
 i \sum_{j=1}^n h_jg_j= i\delta_{\bfal}\hat{\bmath 1}$, coinciding with the result in Eq.(\ref{delta-bfal}). 
 Because the matrix $\Mmual=\bmath 1$, the original and mirrored non-local
 position-type observable coincide, {\it i.e.} $\hat u_{\bfal}=\hat \mu_{\bfal}$.
 The coefficient
 of the mirrored non-local observable $\hat v_{\bfal}$ in Eq.(\ref{old-operators}) 
 can be obtained from Eq.(\ref{hj-gj}) giving the result in Eq.\eqref{Mp-k-row-text}.
 Therefore, the commutator between the mirrored observables is $[\hat u_{\bfal},\hat v_{\bfal}]=i\gamma_{\bfal}\hat{\bmath 1}$, with $\gamma_{\bfal}$
 given in Eq.\eqref{gamma-bfal}.
 \par
In this way,  we see that the coefficients of the mirrored  operators $\hat u_{\bfal}$ and $\hat v_{\bfal}$ are determined once we specify the p-matrix of the bipartition $\Mpal$, whose $k^{\mathrm{th}}$ row
is given in Eq.(\ref{Mp-k-row-text}) and with the  $k^{\mathrm{th}}$ row of $(\Mpal^{-1})^T$ given
in Eq.(\ref{hj}).
Without loss of generality we can choose $k=1$. In 
Appendix \ref{Appendix2}  we give the general structure of a matrix $\Mpal$ with these properties. 
This proves the equality 
$F[\hat \rho^{T_{\bfal}},P_{\hat u_{\bfal}},P_{\hat v_{\bfal}}]=F[\hat \rho,P_{\hat \mu_{\bfal}},P_{\hat \nu_{\bfal}}]$ in Eq.(\ref{criterion-general3-rewrite}).
\par

\section{}
\label{Appendix2}

Here we give the general structure of a $n\times n$ real matrix $\Mpal$ that satisfy Eqs.(\ref{Mp-k-row-text})
and (\ref{hj-gj-a}) for a  given value of $\gamma_{\bfal}$. Without lost of generality we choose $k=1$ so:
\begin{widetext}
\beq
\label{Mpal-matrix-form}
(\Mpal)_{ij}=   
\left\{\begin{matrix} 
      \bar g_1=\bar g_1(\gamma_{\bfal},\delta_{\bfal},\bar g_2,\ldots,\bar g_n,h_1,\ldots,h_n) & \mbox{for  $i=j=1$} \\
      \bar g_j &   \mbox{for  $i=1$ and $1 < j < n$} \\
       \bar g_n=\bar g_n(\gamma_{\bfal},\delta_{\bfal},\bar g_2,\ldots,\bar g_n,h_1,\ldots,h_n) & \mbox{for  $i=1$ and $j=n$} \\
     Q_{ij} & \mbox{ $1<i \leq n$ and $1\leq j \leq n-1$}\\
     -\frac{1}{h_n}\left(\sum_{l=1}^{n-1}h_l Q_{il}\right)& \mbox{for $1< i \leq n$ and $j=n$}
   \end{matrix}\right.
\eeq
\end{widetext}
where  $\bar g_j=-g_{j}$ if  $j$ is one component of the vector $\bfal$ or  $\bar g_j=g_j$ otherwise ($\bfal=\vec \alpha$ or 
$\bfal=\vec \beta$), $\bar g_1$ and $\bar g_n$ are solutions of the equations:
\bea
\sum_{j=1}^n \bar g_j h_j&=&\gamma_{\bfal}\\
\sum_{j=1}^n  g_j h_j&=&\delta_{\bfal}
\eea
and the matrix elements $Q_{ij}$ ($1<i \leq n$ and $1\leq j \leq n-1$) are arbitrary.
 In the case of a ``seed'' partition 
defined by the vector $\vec t =\vec s=(1,\ldots,n_A)$ the explicit form of $\Mpbe$
is:
\begin{widetext}
\beq
\label{seed-matrix-form}
(\Mpbe)_{ij}=   
\left\{\begin{matrix} 
      -g_1=\frac{-1}{h_1}\left(\frac{\delta_{\bfbe}-\gamma_{\bfbe}}{2}-\sum_{l=2}^{n_A}
      g_j h_l\right) & \mbox{for  $i=j=1$} \\
      -g_j &   \mbox{for  $i=1$ and $1\leq j \leq n_A$} \\
      g_j&  \mbox{for  $i=1$ and $n_A <j < n$}\\
      g_n=\frac{1}{h_n}\left(\frac{\delta_{\bfbe}+\gamma_{\bfbe}}{2}-\sum_{l=n_A+1}^{n-1}
      g_j h_l\right) 
      &  \mbox{for  $i=1$ and $j=n$}\\
     Q_{ij} & \mbox{ $1<i \leq n$ and $1\leq j \leq n-1$}\\
     -\frac{1}{h_n}\left(\sum_{l=1}^{n-1}h_l Q_{il}\right)& \mbox{for $1< i \leq n$ and $j=n$}
   \end{matrix}\right. .
\eeq
\end{widetext}

\section{}
\label{Appendix3}

Let's see the concavity of the functional $F_{E}[\hat \rho,P_{\hat \mu},P_{\hat \nu}] = h[P_{\hat \mu}]+ h[P_{\hat {\nu}}]$, {\it i.e.} if $\hat \rho=\sum_j p_j \hat \rho_j$ ($\sum_j p_j=1$) then
$F_{E}[\hat \rho,P_{\hat \mu},P_{\hat \nu}] \geq \sum_j p_j 
F_{j,E}[\hat \rho,P_{\hat \mu},P_{\hat \nu}] $. The Wigner function of  an arbitrary $n$-mode state $\hat \rho$ is
$W(\bsy {x},\bsy {p})=W(\Mxal^{-1}  \vectg{\mu}_{\bfal}, \Mpal^{-1} \vectg{\nu}_{\bfal})$, so
the functional $F_{E}$ can be determined from the marginal distribution (see Eq.(\ref{demostration})):
$\tilde W(\mu,\nu)=\int 
\frac{d{\vectg{{\mu}}_{\bfal}} d{\vectg{{\nu}}}_{\bfal}}{|\gamma_{\bfal}|^{n}}\, 
W(\Mxal^{-1}  \vectg{\mu}_{\bfal}, \Mpal^{-1} \vectg{\nu}_{\bfal})$. If we set
 $\hat \rho=\sum_j p_j \hat \rho_j$ ($\sum_j p_j=1$), the Wigner function of $\hat \rho$ is the convex sum 
 of the Wigner functions of the states $\hat \rho_j$, {\it i.e.} $W(\bsy {x},\bsy {p})=
 \sum_j p_jW_j(\bsy {x},\bsy {p})$,  and therefore for the marginal distribution we have:
 \beq
 \label{App2-marginal}
 \tilde W(\mu,\nu)=\sum_j p_j   \tilde W_j(\mu,\nu).
 \eeq
 Because $P_{\hat {\xi}}(\xi)=\int d\xi^\prime \; \tilde W(\mu,\nu)$ (where $\hat \xi=\hat \nu$ if $\xi^\prime=\nu$ and $\hat \xi=\hat \mu$ if $\xi^\prime=\mu$), from Eq.(\ref{App2-marginal}) we immediately obtain:
 \beq
 P_{\hat {\xi}}(\xi)=\sum_j p_j   P_{j,\hat {\xi}}(\xi).
 \eeq
 Now, we use that the Shannon entropy: $G[P(\xi)]\equiv - \int_{-\infty}^{\infty} d\xi P(\xi) \ln(P(\xi)) $ is a strictly concave functional of $P(\xi)$, {\it i.e.} if $P(\xi)=\sum_{j} p_j P_j(\xi)$
 then $G[P(\xi)]\geq \sum_j p_j G[P_{j,\hat {\xi}}(\xi)]$ (see below), so we can write,
 \bea
 F_{E}[\hat \rho,P_{\hat \mu},P_{\hat \nu}]&=&G[P_{\hat {\mu}}(\mu)]+G[P_{\hat {\nu}}(\nu)]\geq\nonumber\\
 &\geq&  \sum_j p_j \left(G[P_{j,\hat {\mu}}(\mu)]+G[P_{j,\hat {\nu}}(\nu)]\right)=\nonumber\\
 &=&\sum p_j F_{j,E}[\hat \rho,P_{\hat \mu},P_{\hat \nu}] .
 \eea
We can see that $G[P(\xi)]$ is a strictly concave functional in the following way.
First, we note that $P(\xi)=\sum_{j} p_j P_j(\xi)\geq P_{j}(\xi)$, thus because $-\ln(x)$ is a strictly
crescent function we also have $-\ln(P(\xi)) > -\ln(P_{m}(\xi))$. Therefore,
we  we immediately have:
\bea
G[P(\xi)]&=& - \int_S d\xi \left(\sum_{j} p_jP_j(\xi)\right) 
\ln\left(P(\xi)
\right)\geq\nonumber\\
 &\geq&
\sum_{j} p_{j}
\left( - \int_S d\xi P_j(\xi) 
\ln\left(
P_{j}(\xi)
\right)\right)=\nonumber\\
&=&\sum_{j} p_j G[P_j(\xi)].
\eea
 
\bibliographystyle{apsrev}

\end{document}